\documentclass{article}

\usepackage{graphicx}
\usepackage{amssymb}

\title{Radiative transitions in mesons in a non relativistic
quark model}

\author{R. Bonnaz\thanks{{\it E-mail address:} 
 bonnaz@isn.in2p3.fr}, B. Silvestre-Brac\thanks{{\it E-mail address:} 
 silvestre@isn.in2p3.fr; corresponding author}, C. Gignoux\thanks{{\it E-mail address:} 
 gignoux@isn.in2p3.fr} \\
Institut des Sciences Nucl\'eaires, 53 Av. des Martyrs,\\
F-38026 Grenoble Cedex, France.\\}

\begin{document}

\maketitle

\begin{abstract}
In the framework of the non relativistic quark model, an exhaustive
study of radiative transitions in mesons is performed. The emphasis
is put on several points. Some traditional approximations (long wave
length limit, non relativistic phase space, dipole approximation for
E1 transitions, gaussian wave functions) are analyzed in detail and
their effects commented. A complete treatment using three different
types of realistic quark-antiquark potential is made. The overall
agreement with experimental data is quite good, but some improvements
are suggested.

\end{abstract}

\section{Introduction}

\indent

Quantum Chromodynamics (QCD) is today the only reliable theory for
describing strong interactions. There exist many systems that can be
used as a laboratory for exploring and testing the properties of this
basic theory. Among them, the meson and baryon sectors have deserved a
lot of investigations, essentially because they are very easily
produced. However they belong to the non perturbative application of
QCD and thus are not easy to be described from first principles. Despite
many improvements in recent years, on both theoretical and
computational sides, the lattice gauge calculations are still not
completely reliable and cannot explain the whole bulk of known
properties, even for the simplest systems such as the mesons, which
consist of a valence quark-antiquark pair.

This explains why so many various phenomenological approaches have
been developped in order to describe the non perturbative part of QCD.
Among them the non relativistic quark model (NRQM) has met with an
impressive number of successes \cite{Isg78}. The puzzling question is that it still
works even in situations where it is expected to fail; there exist
certainly some deep reasons for such a behaviour although they have
not yet been clarified precisely (see \cite{Sem92}). Basically the NRQM needs to
solve a Schr\"{o}dinger type equation with two-body quark-quark 
(or quark-antiquark) interactions. In recent years, the determination of the
interaction between constituent quarks has reached a high degree of
sophistication and the whole spectra of mesons for instance can be
accounted for in a rather satisfactory way \cite{Sem97}.

However one does know that the description of the spectra is a
necessary but not a sufficient condition for aiming at a good
explanation of non perturbative QCD. In particular several very
different potentials can give rise to spectra of the same quality. One
needs other observables in order to test more precisely the resulting
wave functions. A possibility is the study of static properties, such
as magnetic moments or charged square radii. But more sensitive
observables concern the transitions between various states or
production mechanisms (which rely essentially on the same dynamical
operators). One can think for instance on meson decays under strong
forces (a resonance giving two or several mesons) or the decays under
electroweak forces (a resonance producing a photon or leptons in the
final channel). The advantage of this last kind of transition is that the
transition operator is known perfectly well and thus it is easier to
disantangle the drawbacks coming from less well known strong interactions
through the meson wave function.

In fact this statement is not completely true in the NRQM. Being a
phenomenological theory, NRQM deals with effective degrees of freedom,
the constituent quarks, and a pure Dirac form of the quark-photon 
vertex  in the transition operator is questionnable. Moreover, even in
the traditional approach of radiative transitions (decay of a
resonance into a resonance of lower energy plus a real photon) several
types of approximations are of current use; the effect of these
approximations can hide the necessity of using a more sophisticated
vertex for the quark-photon coupling. Most of those approximations
originate from the formulae widely used in atomic or nuclear physics
which are simply translated in the meson sector. Let us mention the
dipole approximation for E1 transitions, the long wave length
approximation (LWLA) and a non relativistic phase space factor.

Although these approximations are fully justified in atomic or nuclear
physics, it is not obvious that they continue to work when applied to
mesons. Indeed in this sector the transition energy is typically
$E_\gamma$ = $k_\gamma$ = 0.1--0.5 GeV, while the size of the source
is roughly $R$ = 0.5--1 fm = 2.5--5 ${\rm GeV}^{-1}$, so that the long
wave length condition $k_\gamma R << 1$ is not really justified.
Comparing the photon energy to the mass of the emitting meson also
convinces us that a non relativistic phase space is probably not
suited. Moreover the fact that the electrons in an atom or the
nucleons in a nucleus have the same mass is not true in the case of
some mesons, and new phenomena can appear.

During the seventies and eighties, a lot of works has been done on
radiative transitions for mesons (and also for baryons but we are
not so interested in this sector here). At the very beginning they 
were studied in the vector dominance model \cite{Sin70,Bro77}, then
the quark model was introduced either in the framework of MIT bag
model \cite{Hay76,Hac78,Cha84}, the non relativistic quark model
\cite{Odo81,Bar76,Eic78}, 2 body Dirac equation \cite{Mcc83,Bar92,Bar98}
or some relativized phenemenological quark models \cite{God85,Ell93}.
But even in the most complete and nice works, as \cite{God85} or
\cite{Eic78}, there is always an approximation or an inconsistency
which plagues the results or forbids to draw precise conclusions.
Most of relativistic models suffer of a bad treatment of the center
of mass, relativized model do not treat with equal care the quark-quark 
potential and the electromagnetic operator. Moreover in many
papers, a non relativistic phase space or a long wave length
approximation are used and we will see that this is not justified
in the meson sector. In addition, only very few studies concern
the totality of the known experimental data but focus on very
specific transitions (light quark sector or heavy quark sector or
even more restricted sets).

The aim of this paper is essentially twofold. First, within a very
precise framework namely the NRQM, we want to investigate 
deeply and with a particular care the effects of all these
approximations as compared to an exact treatment. Second, we compare,
on an exhaustive list of transitions and using an exact treatment,
how different meson wave functions (obtained with different potentials
giving meson spectra of similar quality) influence the results at the
level of radiative transitions. We have in mind to see whether it is
necessary to modify the quark-photon vertex; our study is thus a first
necessary step before undertaking a more difficult and ambitious program.

In this paper we will consider all the radiative transitions (which
are sufficiently reliable) that are reported in the particle data
group booklet because we want an exhaustive analysis. The experimental
data can be gathered into several groups :

\begin{itemize}
\item the transitions allowed by LWLA; they are essentially M1
transitions ($^3S_1 \rightarrow \: ^1S_0$ or $^1S_0 \rightarrow \: ^3S_1$)
and E1 transitions ($^3P_J \rightarrow \: ^3S_1$ or
$^3S_1 \rightarrow \: ^3P_J$); there is also the particular E1 transition
corresponding to the decay $b_1(1235) \rightarrow \pi \gamma$ (
$^1P_1 \rightarrow \: ^1S_0$)
\item the transitions forbidden by LWLA; they are scarce but
interesting : they correspond to $^3P_J \rightarrow \: ^1S_0$,
$^3S_1 \rightarrow \: ^3S_1$ and $^1P_1 \rightarrow \: ^3S_1$ transitions.
\end{itemize}

The paper is organized as follows. In the next section we present how
the meson wave functions are obtained and also the different 
quark-antiquark potentials that we are studying. In the third chapter we
recall the formalism necessary for the description of radiative
transitions putting the emphasis on the general treatment and the
differences corresponding to the various approximations that we want
to discuss. In the fourth chapter our final expressions for the total
widths are summarized. In the fifth chapter the results are presented and the
effects of each approximation are analyzed in detail. Conclusions are
relegated to the last chapter.

\section{Description of mesons}
\label{sec:mes}

\indent
In the NRQM, the meson is considered as a two particle system : a 
(constituent) quark of mass $m_1$ and an antiquark of mass $m_2$
submitted to a potential $V(r)$, so that the corresponding Schr\"{o}dinger
equation writes :

\begin{equation}
\label{sch}
[m_1+m_2+\frac{{\bf p}^2}{2\mu}+V(r)] \mid \Psi_{\alpha} \rangle = m_{\alpha} \mid
\Psi_{\alpha} \rangle
\end{equation}

\noindent
where $\mu$ is the reduced mass, ${\bf p}$ the relative momentum,
and $m_{\alpha}$ the total mass of the resonance. This last quantity, as well
as the wave function $\mid\Psi_{\alpha} \rangle$, of course depend on the
choice for the potential. The ordinary quarks $u$ and $d$ can be
considered as an isospin doublet; they are noted generically as $n$.
\subsection{The potentials}
\label{ssec:pot}
\indent
In this paper we will consider three different types of potential, the
so-called Bhaduri's (BD) potential \cite{Bha81}, AL1 and AP1 potentials
\cite{Sil93}. For
the purpose of our analysis, it is not necessary to introduce very
sophisticated forms including spin-orbit, tensor forces, ...whose
effects are not so important and which complicate quite a lot the
formalism. We limit ourselves to potentials containing two different
structures : a central term and a hyperfine term; this is the minimum
requirement to get reliable wave functions :

\begin{equation}
\label{pot}
V(r)=V_c(r)+V_h(r) {\bf \sigma_1}.{\bf \sigma_2}
\end{equation}

In this expression, we forget about the color term which is always the
same whatever the meson under consideration.

The central term $V_c(r)$ contains a short range coulomb part (remnant
of one gluon exchange) and a confining term :

\begin{equation}
\label{potc}
V_c(r)= - \frac{\kappa}{r} + a r^p - C
\end{equation}

BD and AL1 potentials exhibit a traditional linear confinement ($p=1$)
while AP1 potential uses $p=2/3$, a power best suited to Regge
trajectories in non relativistic dynamics.
In each potential the hyperfine term is short range (remnant of the
Dirac factor of the Fermi-Breit approximation). For BD it is of Yukawa
type:

\begin{equation}
\label{pothbd}
V_h(r)= \frac{\kappa}{m_1 m_2}\frac{exp(-r/r_0)}{r r_0^2}
\end{equation}

For AL1 and AP1 it is of gaussian type:

\begin{equation}
\label{poth}
V_h(r)= \frac{2 \kappa^{\prime}}{3 m_1 m_2} \frac{exp(-r^2/r_0^2)}{\pi^
{1/2}/ r_0^3}
\end{equation}
\noindent
but, contrary to BD, the size is mass dependent through
\begin{equation}
\label{r0}
r_0 = A (2 \mu)^{-B}
\end{equation}
The parameters for the potentials are gathered in table (\ref{parapot}).
The meson spectra obtained with these potentials are rather good and
have been presented elsewhere \cite{Sil93}. It is important to stress that they
also give quite good results in the baryon sector; so we have some 
confidence that they describe the quark dynamics in a meson in a 
rather satisfactory way.

\begin{table}
\caption{ \label{parapot} Parameters for the three used potentials,
defined by (\ref{pot}-\ref{r0}). All the units are in power of GeV.}
\begin{center}
$\begin{tabular}{|c|c|c|c|}\cline{2-4}
\multicolumn{1}{c|}{} & BD & AL1 & AP1 \\ \hline
$m_u=m_d=m_n$	 & 0.337 & 0.315 & 0.277  \\ \hline
$m_s$		 & 0.600 & 0.577 & 0.553  \\ \hline
$m_c$	         & 1.870 & 1.836 & 1.819  \\ \hline
$m_b$		 & 5.259 & 5.227 & 5.206  \\ \hline
$\kappa$         & 0.520 &0.5069 & 0.4242  \\ \hline
$a$		 & 0.186 &0.1653 & 0.3898  \\ \hline
$p$		 & 1     & 1	 & 2/3  \\ \hline
$C$		 &0.9135 &0.8321 & 1.1313  \\ \hline
$\kappa^{'}$	 & -     &1.8609 & 1.8025  \\ \hline
$r_0$		 & 2.305 & -	 & -  \\ \hline
$A$		 & -     &1.6553 & 1.5296  \\ \hline
$B$		 & -     &0.2204 & 0.3263  \\ \hline

\end{tabular}$
\end{center}
\end{table}

\subsection{Meson wave functions}
\label{ssec:mwv}
Because of the rotational invariance, the meson wave function is written as:
\begin{equation}
\label{mfo}
\mid \Psi_{ILSJ} \rangle = \eta_I(1,2) [\Phi_{nL}(1,2) \chi_S(1,2)]_J
\end{equation}
\noindent
where $\eta_I(1,2)$ is the isospin wave function with total isospin
$I$, $\chi_S(1,2)$ is the spin wave function with total spin $S$ and
$\Phi_{nL}(1,2)$ the space wave function with orbital momentum $L$ and
radial number $n$. Spin $S$ and orbital momentum $L$ are coupled to
total angular momentum $J$, but are nevertheless good quantum numbers.
The color wave function is forgotten since
it does not play any role in the formalism; the flavor content of the
meson is also forgotten since the electromagnetic operator does not
change the quark flavor. The magnetic quantum numbers are not
indicated here although the projection of the isospin plays a role
since the operator is a mixing of isoscalar and isovector terms. We
will come back on this point later on.

\begin{figure}[htb]
\begin{center}
\includegraphics[width=0.8\linewidth]{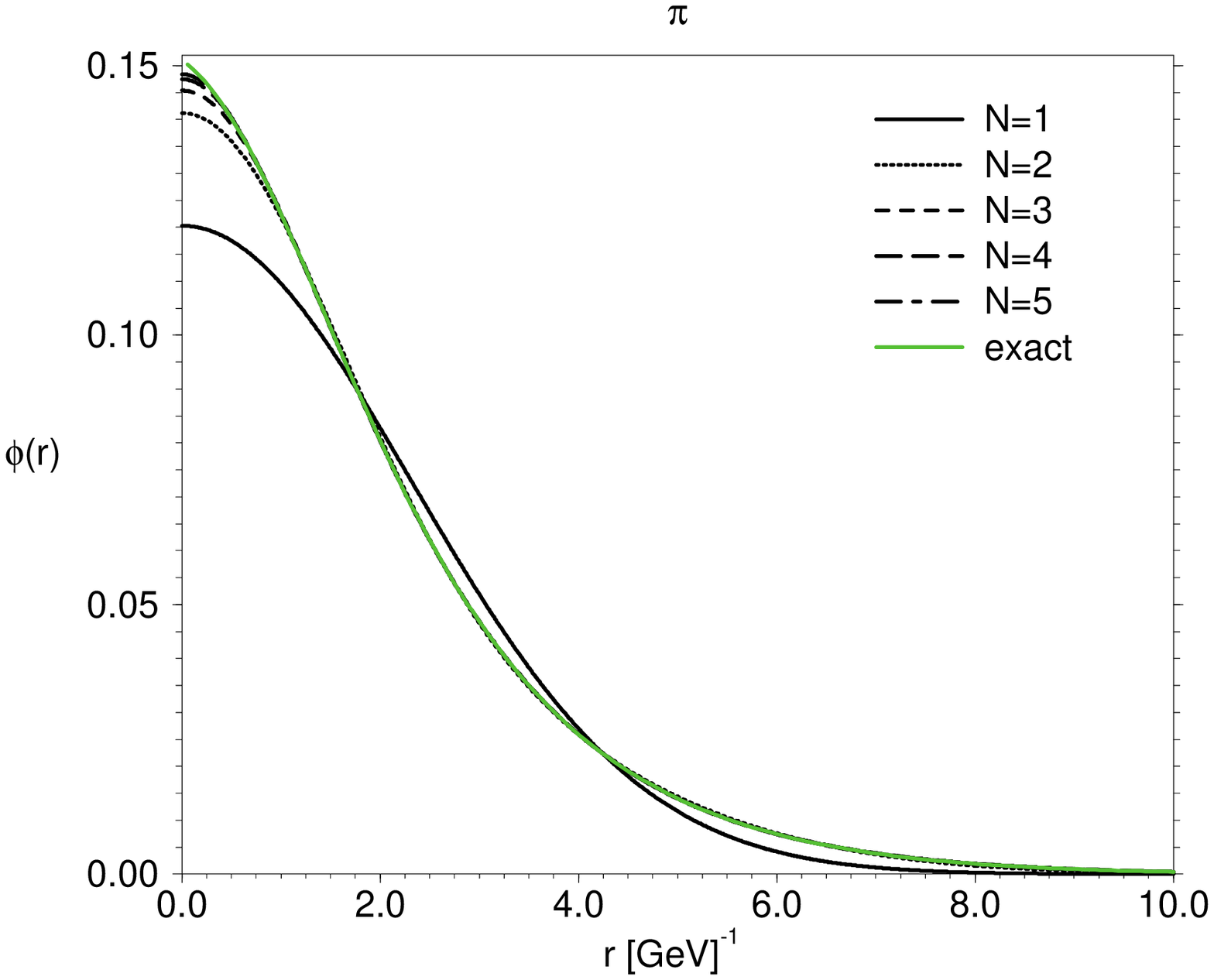}
\includegraphics[width=0.8\linewidth]{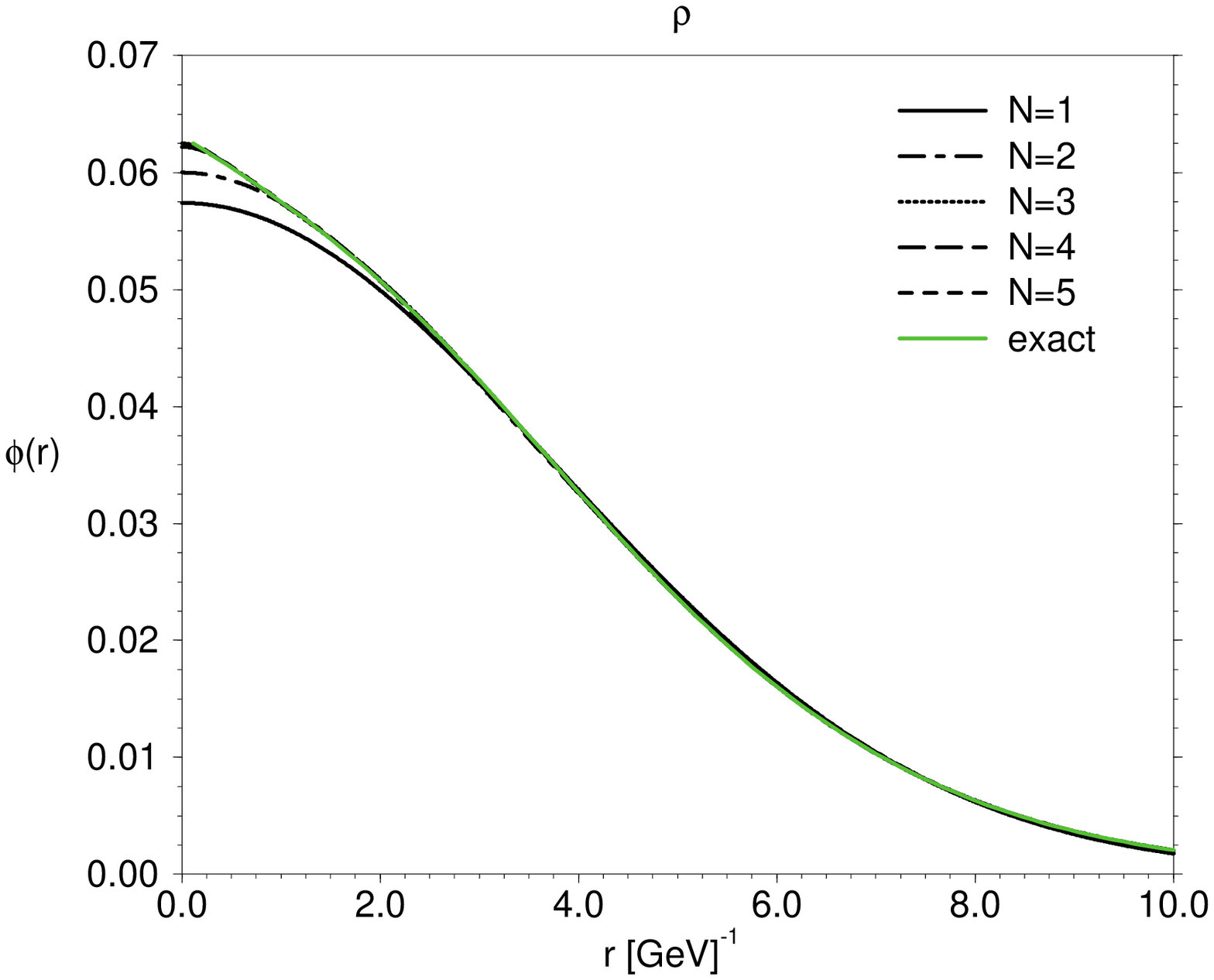}
\caption{\label{fig1} $\pi$ and $\rho$ radial wave functions 
$\frac{R_{nl}(r)}{\sqrt{4\pi}}$ for N=1 to 5 gaussian terms
 compared to the exact ones.}
\end{center}
\end{figure}

We will use the expression of the space wave function both in position
representation and in momentum representation. Thus we write :

\begin{eqnarray}
\label{foe}
\langle {\bf r} \mid \Phi_{nL}(1,2)\rangle & = &  R_{nL}(r) Y_L(\hat{r})
\nonumber \\
\langle {\bf p} \mid \Phi_{nL}(1,2)\rangle& = &  R_{nL}(p) Y_L(\hat{p})
\end{eqnarray}
We use the same notation for the radial wave function $R_{nL}$ in both
representations if there is no risk of confusion.
In fact the momentum representation is very often more convenient to
deal with the matrix elements of the transition operator. Moreover an
approximation of the exact wave function in terms of gaussian
functions will be particularly well suited to compute easily difficult quantities.
In this case we write:

\begin{equation}
\label{fop}
R_{nL}(p)=p^L \sum_{i=1}^{N}c_i \exp(-A_i p^2/2)
\end{equation}
\noindent
and a similar expression for the wave function in coordinate
representation. In general the number of gaussian terms needed
in (\ref{fop}) to achieve convergence is rather weak. Just to give
an idea of the quality of such an expansion in the case of AL1 potential,
the masses $m_{\pi}$ and $m_{\rho}$ as function of $N$ are presented in Table
(\ref{mpiro}) and compared to the exact ones (obtained by solving the
differential equation with the Numerov algorithm). In the same spirit
we plot the corresponding wave functions in figure (\ref{fig1}). Although
there is some differences between the $N$=1 and $N$=2 cases, one sees
that very rapidly the approximation (\ref{fop}) can be identified to the
exact solution. In practice we perform our calculations with $N$=5 and
consider the corresponding wave function as the exact one.
Let us also point out that, since the basis states are not orthogonal,
different sets of ($c_i,A_i$) parameters can be used with equal
success. In fact, we determine these coefficients by two different
methods : i) minimisation of the state with respect to these
parameters, ii) best fit of the exact wave function with a form like
(\ref{fop}). For $N$=5, both procedures give exactly the same results.
Thus in the rest of the paper we qualify the wave function given by
(\ref{fop}) with $N=5$ as the exact wave function.

\begin{table}
\caption{ \label{mpiro} Convergence properties, with three different
quark-antiquark potentials, for the $\pi$ and $\rho$ masses as
function of the number of Gaussian functions NG in (\ref{fop}). This 
exact value is obtained with the Numerov algorithm.}
\begin{center}
$\begin{tabular}{|cc|c|c|c|}\cline{3-5}
\multicolumn{2}{c|}{} & BD & AL1 & AP1 \\ \hline
$\pi$& 1G & 0.252050 & 0.193702 & 0.192776   \\ \hline
  &   2G  & 0.157151 & 0.140359 & 0.140285  \\ \hline
  &   3G  & 0.141707 & 0.138436 & 0.138391  \\ \hline
  &   4G  & 0.138835 & 0.138220 & 0.138192  \\ \hline
  &   5G  & 0.138233 & 0.138188 & 0.138168  \\ \hline
 \multicolumn{2}{|c|}{exact} & 0.138186  & 0.138057 & 0.138044   \\ \hline

$\rho$&1G & 0.780344 & 0.771104 & 0.772196  \\ \hline
  &   2G  & 0.778425 & 0.770731 & 0.770166  \\ \hline
  &   3G  & 0.778402 & 0.770033 & 0.770156  \\ \hline
  &   4G  & 0.778375 & 0.770013 & 0.770036  \\ \hline
  &   5G  & 0.778374 & 0.770006 & 0.770029  \\ \hline
 \multicolumn{2}{|c|}{exact} & 0.778347 & 0.769958 & 0.769988   \\ \hline
\end{tabular}$
\end{center}
\end{table}

\section{Radiative transitions}
\label{sec:trrad}
\indent
A number of points are already well known, and we do not want to spend
too much time on them. We will focus our attention essentially on new
aspects or on formulations that are discussed in details later on.
Everywhere in this paper we employ natural units $\hbar=c=1$.

\subsection{Transition operator}
\label{ssec:oper}

We start with the non relativistic expression of the electromagnetic
transition operator between an initial meson state and a final meson
state plus a real photon of momentum ${\bf k}$, energy
E$ = \mid {\bf k} \mid$ and polarisation ${\bf \epsilon}
({\bf k},\lambda)$. We adopt, as usual, the Coulomb gauge, and we
normalize the plane waves in a quantification volume $V$.

\begin{eqnarray}
\label{hi}
H_I&=&- \sqrt{\frac{2 \pi \alpha}{V E}}{\bf \epsilon}
({\bf k},\lambda) \cdot {\bf M}\\
\label{bigM}
{\bf M}&=&\sum_{i=1}^2 \frac{e_i}{2 m_i}\
\exp (-i {\bf k} \cdot {\bf r}_i) ( 2 {\bf p}_i  -i {\bf \sigma}_i
\times {\bf k})
\end{eqnarray}
The summation runs on the two particles of charge $e_i$ and mass $m_i$
present in the meson. The first term in ${\bf M}$ is known as the electric term and
the second one as the magnetic term.

\subsection{Transition amplitude}
\label{ssec:amp}
\begin{figure}[htb]
\begin{center}
\includegraphics{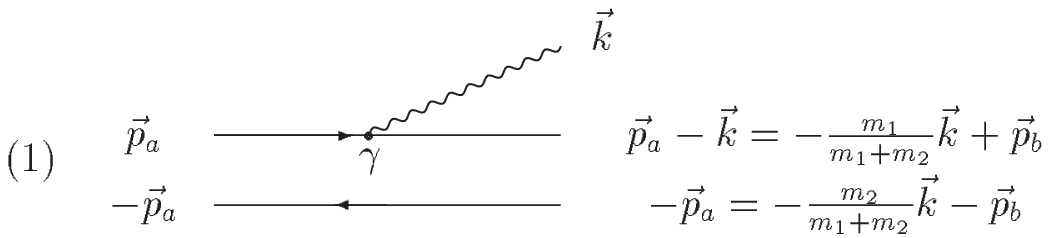}\vspace{2cm}
\includegraphics{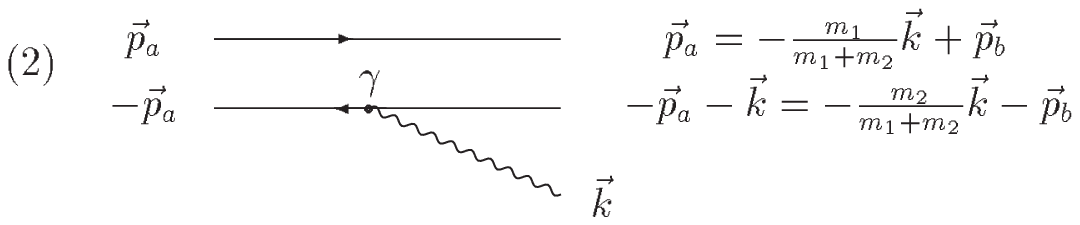}
\caption{\label{schema} \normalsize Schematic representation of the elementary decay process with momentum values from the rest frame of the inital meson. (1) denotes the emission of the real photon by the quark and gives rise to $M^{(1)}$ and (2) by the antiquark and gives rise to $M^{(2)}$.}
\end{center}
\end{figure}

The initial meson of mass $m_a$ is at rest and has angular momentum
$J_a M_a$ (coupling of $L_a$ and $S_a$), isospin $I_a M_{I_a}$. The
final meson of mass $m_b$ has a total momentum ${\bf K}_b$, angular
momentum $J_b M_b$ (coupling of $L_b$ and $S_b$), isospin
$I_b M_{I_b}$. Inserting ${\bf M}$ of (\ref{bigM}) between the
corresponding wave functions (in the rest frame of the meson ${\bf p}_1$ = - ${\bf p}_2$
= ${\bf p}$) in momentum representation (\ref{mfo})
(completed by the center of mass plane wave), one gets the transition
amplitude:

\begin{equation}
\label{tramp}
 {\bf M}_{A \rightarrow  B} = \delta_{{\bf K}_b,-{\bf k}}  
[{\bf M}^{(1)}_{A \rightarrow B} +{\bf M}^{(2)}_{A \rightarrow B}]
\end{equation}

\begin{eqnarray}
\label{amp12}
 {\bf M}^{(1)}_{A \rightarrow B}& = &
\int d^3p \, \Phi^*_B({\bf p}-\frac{m_2}{m_1+m_2} {\bf k})
[2 {\bf p}- i {\bf \sigma}_1 \times {\bf k}]
\frac{<e_1>}{2 m_1} \Phi_A({\bf p}) \\
 {\bf M}^{(2)}_{A \rightarrow B}& = &
\int d^3p \, \Phi^*_B({\bf p}+\frac{m_1}{m_1+m_2} {\bf k})
[- 2 {\bf p}- i {\bf \sigma}_2 \times {\bf k}]
\frac{<e_2>}{2 m_2} \Phi_A({\bf p})
\nonumber 
\end{eqnarray}
\noindent
where the subscripts refer to the particle number, see figure (\ref{schema}).

\large

\normalsize
We must say a word about the charge and its relation to isospin. One
can perfectly well ignore isospin degrees of freedom and speak only of
flavor wave function and its symmetry properties. In this case $e_i$
is simply a physical quantity, the charge corresponding to the flavor
of quark i; we understand $<e_i>$ appearing in (\ref{amp12}) as $e_i$.
On the other hand, one can also introduce isospin degrees of freedom
for convenience, in particular, it makes things easier in the case of 
neutral mesons with flavor mixing or mixing angles; for isospin  $t_i$=0
 one still has $<e_i>$ = $e_i$. However
for isospin $t_i$=1/2, the charge is different for each member of the
multiplet so that it becomes an isospin dependent operator
$e_i= \pm 1/6 +(t_i)_z$ (+ for quark, - for antiquark). In this case on
must understand $<e_i>$ as the matrix element of this operator on
isospin wave function $\eta_I$. If we adopt this philosophy, a very
simple calculus, based on Racah algebra, gives :\\

\begin{eqnarray}
\label{ei}
<e_i> & = & e_i\, \delta_{I_aI_b}\, \delta_{M_{I_a}M_{I_b}} \,\hspace{1.5cm} t_i=0
\nonumber \\
<e_1> &=& \delta_{M_{I_a}M_{I_b}} \left[ \frac{1}{6}\delta_{I_aI_b} +
(-1)^{I_a+t_2-1/2} \sqrt{3(I_a+1/2)}\right.  \nonumber \\ 
&&\left.  <I_a M_{I_a} 1 0 \mid I_b M_{I_b}>
\left\{ \begin{tabular}{ccc}
1 & 1/2 & 1/2 \\
$t_2$ & $I_a$ & $I_b$\\
\end{tabular} \right\}
\right] \,\hspace{0.5cm} t_1=1/2   \\
<e_2> &=& \delta_{M_{I_a}M_{I_b}} \left[- \frac{1}{6}\delta_{I_aI_b} +
(-1)^{I_b+t_1-1/2} \sqrt{3(I_a+1/2)}\right.  \nonumber \\ 
&&\left. <I_a M_{I_a} 1 0 \mid I_b M_{I_b}> 
\left\{\begin{tabular}{ccc}
1 & 1/2 & 1/2 \\
$t_1$ & $I_a$ &$ I_b$ \\
\end{tabular} \right\}
\right] \, \hspace{0.5cm} t_2=1/2  \nonumber
\end{eqnarray}

\subsubsection{Long Wave Length Approximation}
\label{sssec:lwla}
Due to the recoil term in meson $B$ wave function, the space term
appearing in (\ref{amp12}) is not easy to calculate. A widely used
approximation is the long wave length approximation (LWLA) which
consists in putting ${\bf k}$=0 in the argument of $\Phi_B$ (this is
equivalent to replace $\exp(-i{\bf k} \cdot {\bf r})$ by 1 in
coordinate representation).
It is just a matter of Racah algebra to disantangle the spin and space
degrees of freedom in (\ref{amp12}). It is pleasant that, in this case,
the electric and magnetic part, which are of different parity, cannot
couple the same states; this is why we speak about electric and
magnetic transitions. It is also convenient to calculate the covariant
spherical components $\mu = -1,0,1$ of the vector ${\bf M}$.

The electric transitions change the parity of the state, but not the
spin. One has, for the $\mu$ component :

\begin{equation}
\label{Me}
({\bf M}^e_{A \rightarrow B})_{\mu}= \delta_{S_a,S_b} \delta_{{\bf K}_
b,-{\bf k}} < J_a M_a 1 \mu \mid J_b M_b > F^{e}(A \rightarrow B)
\end{equation}
\noindent
with the expression $F^{e}$, independent of the magnetic numbers:

\begin{eqnarray}
\label{fe}
F^{e} (A \rightarrow B) &=& (-1)^{S_a+J_a+L_b+1} (\frac{<e_1>}{m_1}-
\frac{<e_2>}{m_2}) \hat{L_a} \hat{J_a} < L_a 0 1 0 \mid L_b 0 > 
\nonumber \\
&&\left\{ \begin{tabular}{ccc}
$L_a$ & $S_a$ & $J_a$ \\
$J_b$ & 1 & $L_b$\\
\end{tabular}
\right\} 
\int _0^{\infty}  dp \, R_{n_bL_b}(p) R_{n_aL_a}(p) p^3
\end{eqnarray}
\noindent
where we have introduced the usual notation $\hat{J}$=$\sqrt{2J+1}$.

The magnetic transitions do not change parity nor orbital momentum.
Since the operator contains a spin dependent part, the decoupling is a
little bit more complicated but straightforward anyhow:

\begin{eqnarray}
\label{Mm}
({\bf M}^m_{A \rightarrow B})_{\mu}&=& \delta_{L_a,L_b} \delta_{{\bf K}_
b,-{\bf k}}  F^{m}(A \rightarrow B)  \nonumber \\
&&\sum_{\nu,\sigma}
< 1 \nu 1 \sigma \mid 1 \mu > < J_a M_a 1 \nu \mid J_b M_b >
k_{\sigma}
\end{eqnarray}
\noindent

with the expression $F^{m}$, independent of the magnetic numbers:

\begin{eqnarray}
\label{fm}
F^{m}(A \rightarrow B)&=& (-1)^{L_a+J_b} \sqrt{3} (\frac{<e_1>}{m_1}+
(-1)^{S_a+S_b} \frac{<e_2>}{m_2}) \hat{S_a} \hat{J_a} \hat{S_b} 
\nonumber \\
&& \left\{ \begin{tabular}{ccc}
$S_a$ & $L_a$ & $J_a$ \\
$J_b$ & 1 & $S_b$ \\
\end{tabular}
\right\}  \left\{
\begin{tabular}{ccc}
 1 & 1/2 & 1/2 \\
1/2 & $S_a$ & $S_b$ \\
\end{tabular}
\right\} \\
&& \int_0^{\infty}dp \, R_{n_bL_b}(p) R_{n_aL_a}(p) p^2 \nonumber
\end{eqnarray}

Of course, when the gaussian expansion of the wave function (\ref{fop})
is employed, the dynamical integrals are analytical.

\subsubsection{Beyond long wave length approximation}
\label{sssec:}
The big advantage of using the wave functions expressed on gaussian
terms (\ref{fop}) is that the treatment of the general case can be dealt
with rather simply. Since taking $N$=5 is equivalent to treat the
exact wave function, the following treatment solves exactly the
problem. In fact an individual term in the expansion is of the form
$\Phi_L({\bf p})=\exp(-A p^2/2) \mathcal{Y}_L({\bf p})$ where
$\mathcal{Y}_L({\bf p})=p^L Y_L(\hat{p})$ is the usual solid harmonic.
Just to illustrate the procedure, let us consider only one term in the
expansion (one $A$ quantity for meson A, one $B$ quantity for meson B)
with a unit amplitude $c^A=c^B=1$). The argument in meson B wave
function is now a linear combination of ${\bf p}$ and ${\bf k}$. Such
a combination in the solid harmonic can be treated using the formulae
:

\begin{eqnarray}
\label{cby}
\mathcal{Y}_{lm}(a {\bf p}_1+b{\bf p}_2)&=&\sum_{l_1=0}^l
C_{l_1}^l a^{l_1} b^{l-l_1} [\mathcal{Y}_{l_1}({\bf p}_1)
\mathcal{Y}_{l-l_1}({\bf p}_2)]_{lm} \nonumber \\
C_{l_1}^l&=&\sqrt{\frac{4 \pi (2l+1)!}{(2 l_1+1)!(2(l-l_1)+1)!}}
\end{eqnarray}
and
\begin{eqnarray}
\label{cpy}
[\mathcal{Y}_{l_1}({\bf p})\mathcal{Y}_{l_2}({\bf p})]_{lm}&=&
B_{l_1 l_2}^l p^{l_1+l_2-l} \mathcal{Y}_{lm}({\bf p}) \nonumber \\
B_{l_1 l_2}^l&=&(-1)^l \frac{\hat{l_1}\hat{l_2}}{\sqrt{4 \pi}}
\left(
\begin{tabular}{ccc}
$l_1$ & $l_2$ & l \\
0 & 0 & 0
\end{tabular}
\right)
\end{eqnarray}

The same combination appearing in the exponential is a quadratic form
which can be diagonalized in order to get rid of the non diagonal
terms. Let us just summarize our conclusions.

Here are some auxilliary quantities ($i=1,2$ refers to particle number):
\begin{equation}
\label{xyd}
D=\frac{A+B}{2} \; ; \; x^{(i)}=\frac{m_{3-i}B}{(m_1+m_2)(A+B)} 
\end{equation}
%\begin{equation}\nonumber
$$z^{(i)}=\frac{m_{3-i}A}{(m_1+m_2)(A+B)} \; ; \; F^{(i)}=D x^{(i)} z^{(i)}$$
%\end{equation}
If the masses of the quark and the antiquark are the same there is no
need to distinguish the quantities $x,z,F$ and further simplifications
arise which we do not want to comment. The transition amplitudes in
the general case can be converted to a more appropriate form:

\begin{eqnarray}
\label{lwm}
{\bf M}^{(1)}_{A \rightarrow B}&=&\frac{<e_1>}{2 m_1}
\int d^3q \exp(-D q^2 - F^{(1)} k^2) [\mathcal{Y}^*_{L_b}
({\bf q}-z^{(1)}{\bf k})\chi_{S_b}]_{J_b}\nonumber \\
&&[2 {\bf q} - i {\bf \sigma}_1 \times {\bf k}]
[\mathcal{Y}_{L_a}({\bf q}+x^{(1)}{\bf k})\chi_{S_a}]_{J_a} \nonumber
\\
{\bf M}^{(2)}_{A \rightarrow B}&=&\frac{<e_2>}{2 m_2}
\int d^3q \exp(-D q^2 - F^{(2)} k^2) [\mathcal{Y}^*_{L_b}
({\bf q}+z^{(2)}{\bf k})\chi_{S_b}]_{J_b}\nonumber \\
&&[-2 {\bf q} - i {\bf \sigma}_2 \times {\bf k}]
[\mathcal{Y}_{L_a}({\bf q}-x^{(2)}{\bf k})\chi_{S_a}]_{J_a}
\end{eqnarray}

Altough it is possible to pursue the calculations using (\ref{lwm}),
experimentally all the known transitions exhibit either $L_a$=0 or
$L_b$=0. The resulting formulae look much more sympathetic in those
cases and we report here only these special cases since only them will
be applied in the next chapter.
One important difference, as compared to the LWLA, is that both the
electric and magnetic terms of the operator contributes to a given
transition. Let us present the result for $L_b=0$.

\begin{equation}
\label{trla}
({\bf M}_{A \rightarrow BL_b=0})_{\mu}=\delta_{S_b,J_b}
[\mathcal{E}_{\mu}(A;x)+\mathcal{M}_{\mu}(A;x)]
\end{equation}

The term $\mathcal{E}_{\mu}(A;x)$ comes from the electric part of the
operator and writes:

\begin{equation}
\label{emu}
\mathcal{E}_{\mu}(A;x)=\delta_{S_a,S_b} \frac{\sqrt{4 \pi L_a}}{6}
\hat{L_a} \Gamma(5/2)
 < L_a, M_a-M_b; J_b, M_b \mid J_a, M_a >
 \end{equation}
$$< 1, -\mu; L_a-1, M_a-M_b+\mu \mid L_a, M_a-M_b >
(-1)^{\mu}\mathcal{Y}_{L_a-1,M_a-M_b+\mu}({\bf k})$$
$$\left[
\frac{<e_1>}{m_1} \frac{(x^{(1)})^{L_a-1} \exp(-F^{(1)}k^2)}{D^{5/2}}
+(-1)^{L_a} \frac{<e_2>}{m_2} \frac{(x^{(2)})^{L_a-1}
\exp(-F^{(2)}k^2)}{D^{5/2}} \right]$$
\vspace{0.5cm}

Note that this term vanishes if $L_a=0$.

The term $\mathcal{M}_{\mu}(A;x)$ comes from the magnetic part of the
operator and writes:

\begin{equation}
\label{mmu}
\mathcal{M}_{\mu}(A;x)=(-1)^{1+S_a} 2 \pi \Gamma(3/2) \hat{S_a} \left\{
\begin{tabular}{ccc}
1 & 1/2 & 1/2 \\
1/2 & $S_a$ & $J_b$\\
\end{tabular}
\right\}
\end{equation}
$$\left[ \frac{<e_1>}{m_1} \frac{x^{(1)\ L_a}
\exp(-F^{(1)}k^2)}{D^{3/2}}+(-1)^{L_a+S_a+J_b} \frac{<e_2>}{m_2} \frac
{x^{(2)\ L_a} \exp(-F^{(2)}k^2)}{D^{3/2}} \right] $$
$$ \sum_{\mu_a,\sigma_a,\nu, \sigma} < L_a \mu_a S_a \sigma_a
\mid J_a M_a > < S_a \sigma_a 1 \nu \mid J_b M_b > < 1 \nu 1 \sigma
\mid 1 \mu > \mathcal{Y}_{1 \sigma}({\bf k}) \mathcal{Y}_{L_a \mu_a}({\bf k})$$
\vspace{0.5cm}

The case $L_a=0$ looks very similar, but one has to be very careful
with the phases. In this case the electric part is given by

\begin{equation}
\label{emu2}
\mathcal{E}_{\mu}(B;z)=(-1)^{L_b-1-\mu} \mathcal{E}^*_{-\mu}(A=B;x=z)
\end{equation}
meaning that in expression (\ref{emu}), one has to change all quantities
relative to A by the corresponding ones relative to B, change the sign
of $\mu$, change $x$ by $z$, take the complex conjugate and multiply
by a given phase. The expression for  $\mathcal{M}_{\mu}(B;z)$ is
obtained with the same prescription as (\ref{emu2}).

If one admits more than one gaussian function in the expansion of the wave
function the quantities defined in (\ref{xyd}) depends on which terms
are retained and must be written more explicitly i.e $D_{ij}$ =
$(A_i+B_j)/2$. To obtain the complete expression corresponding to the
exact wave function, one must take care of this; for example one must
make in (\ref{emu}-\ref{mmu}) the following replacement:

\begin{equation}
\label{rep}
\frac{(x^{(1)})^{L_a-1} \exp(-F^{(1)}k^2)}{D^{5/2}} \rightarrow
\sum_{i,j=1}^{N} c_i^A c_j^B \frac{(x_{ij}^{(1)})^{ L_a-1}
\exp(-F_{ij}^{(1)}k^2)}{D_{ij}^{5/2}}
\end{equation}

and similar replacements everywhere.

\section{The phase space}

\subsection{density of states}
The density of states is obtained with periodic conditions in the quantification
box. The treatment can be found in any textbook. The density of states by
energy unit and by solid angle unit is given by
\begin{equation}
\label{den}
\rho(E,\Omega) = \frac{VE^2}{(2\pi)^3}
\end{equation}
One has to calculate the matrix element $<B\gamma|H_i|A>^2$ from (\ref{hi}). The
best way  is to use spherical components for the vectors and Racah algebra to 
deal with the corresponding expressions. One has then to sum over the 
polarisations of the photon and of the final meson and to average over the 
polarisations of the decaying meson. In order to simplify the notations, let
us introduce the quantity $X(E)$ by:
\begin{equation}
\label{xe}
X(E)\delta_{\vec{K}_b,-\vec{k}}=\frac{1}{\hat{J_A}^2}\sum_{\lambda=\pm1}
\sum_{M_a,M_b} \mid <B\gamma|H_i|A> \mid ^2
\end{equation} 
The decay width is given by the golden rule:
\begin{equation}
\label{gt}
\Gamma = \int dE\ \delta(E_f-E_i)\int d\Omega\ 2\pi\ \rho(E, \Omega) X(E)
\end{equation}
In the litterature, one finds different formulae for the width depending upon
how is treated the energy Dirac term in (\ref{gt}). This is known as the
phase space factor $\Phi$.
Explicitly, one writes :
\begin{equation}
\label{gpt}
\Gamma = \Phi(E_0) V X(E_0)
\end{equation}
As this must be, the quantification volume $V$, appearing in (\ref{gpt}), 
cancels with the one present in $X(E)$ as seen in (\ref{hi}). $E_0$ denotes
the energy value fulfilling the equation $E_f=E_i$.

\subsection{relativistic phase space}
In this case the energies $E_f$ and $E_i$ are given by their relativistic
expressions. Taking into account the fact that the momentum of the final
meson is opposite to the photon momentum, the Dirac factor is 
$\delta(E_B+E-m_a)=\delta(\sqrt{m_b^2+E^2}+E-m_a)$. The integral in (\ref{gt})
is performed with the usual rules on delta functions to give :
\begin{equation}
\label{pr}
\Phi(E_0)= \frac{E_0^2}{\pi} \frac{E_B(E_0)}{m_a} \; ; \;
E_0 = \frac{m_a^2-m_b^2}{2m_a}
\end{equation}

\subsection{Non relativistic phase space}
In a non relativistic treatment, the energies are related to their momenta by
the classical expressions. Alternatively, one can make the approximation
$ E_0<< m_a,\ m_b$ in the relativistic phase (\ref{pr}).
\begin{equation}
\label{pnr}
\Phi(E_0)= \frac{E_0^2}{\pi}  \; ; \; E_0 = m_a-m_b
\end{equation}
Thus, as compared to the relativistic expression, the non relativistic phase
space differs by two effects. The energy of the photon is equal to the
energy difference between the resonances (as in nuclear physics) and the term
$\frac{E_B(E_0)}{m_a}$ is equal to unity. 

\subsection{mixed phase space}
The mixed phase space is sometimes used in litterature \cite{Bar92}. It consists
in retaining the relativistic value (\ref{pr}) for the energy $E_0$ -- because
the effect should be important in the meson sector -- but the non relativistic
value (\ref{pnr}) for $\Phi(E_0)$. Thus the mixed phase space is based on :
\begin{equation}
\label{pmix}
\Phi(E_0)= \frac{E_0^2}{\pi}  \; ; \; E_0 = \frac{m_a^2-m_b^2}{2m_a}
\end{equation}

\section{Total widths}
\subsection{long wave length approximation}
\label{ssec:twlwla}
With expression (\ref{Me}) for the amplitude and (\ref{gpt}) and (\ref{pr}) for the
phase space, the total width for electric transition is obtained after some
calculations:
\begin{eqnarray}
\label{lte}
\Gamma^e_{A \rightarrow B \gamma}&=&\delta_{S_a,S_b} 4 \frac{\alpha}{3} E_0
\frac{E_B(E_0)}{m_a} (\frac{<e_1>}{m_1}-\frac{<e_2>}{m_2})^2 \nonumber \\
&&\hat{L_a}^2 \hat{J_b}^2 < L_a 0 1 0 \mid L_b 0 >^2  
\left\{ \begin{tabular}{ccc} 
$L_{a}$ & $S_{a}$ & $J_{a}$ \\
$J_{b}$ & 1 & $L_{b}$\\
\end{tabular}
\right\}^2  \nonumber \\
&& \left[ \int _0^{\infty}dp \, R_{n_bL_b}(p) R_{n_aL_a}(p) p^3 \right]^2
\end{eqnarray}
In the same way the magnetic transition width results from the expression
(\ref{Mm}) for the amplitude. It looks like :
\begin{eqnarray}
\label{ltm}
\Gamma^m_{A \rightarrow B \gamma}&=&\delta_{L_a,L_b} 2 \alpha E_0^3 \frac{E_B(E_0)}{m_a}
(\frac{<e_1>}{m_1}+(-1)^{S_a+S_b}\frac{<e_2>}{m_2})^2 \nonumber \\
&&\hat{S_a}^2 \hat{S_b}^2 \hat{J_b}^2 
\left\{ \begin{tabular}{ccc} 
1 & 1/2 & 1/2 \\
1/2 & $S_a$ & $S_{b}$\\
\end{tabular}
\right\}^2 
\left\{ \begin{tabular}{ccc} 
$S_{a}$ & $L_{a}$ & $J_{a}$ \\
$J_{b}$ & 1 & $S_{b}$\\
\end{tabular}
\right\}^2  \nonumber \\
&& \left[ \int _0^{\infty}dp \, R_{n_bL_b}(p) R_{n_aL_a}(p) p^2 \right]^2
\end{eqnarray}
These formulae hold for a relativistic phase space. The modification due to the
use of a non relativistic or a mixed phase space results obviously from the
discussion of the previous chapter.
Simplified expressions applied to specific transitions seen experimentally are
relegated to the appendix.\\
Very often, the electric transition is calculated using the dipole approximation.
The advantage is that it is not necessary to have the meson wave function in
momentum representation but in coordinate representation which is a more natural
scheme, while avoiding a complicated nabla operator. However one must be very 
careful in handling it because it relies on some
approximations which are not always justified. Indeed the dynamical factor
appearing in the amplitude is proportionnal to :
$\langle B\mid {\bf p} \mid A\rangle$. The widely used trick is to remark that
${\bf p}$ is itself proportionnal to $\left[\ H_{int},\ {\bf r}\ \right]$ so
that $\langle B\mid {\bf p} \mid A\rangle\ = i \mu (m_a - m_b)
 \langle B\mid {\bf r} \mid A\rangle $, where $\mu$ is the reduced mass of the
 meson system (this is sometimes known as the Siegert theorem).

\noindent Thus the only change to the previous derivation of the decay width for
 the eletric term in equation (\ref{lte}) is the replacement of
$ \int _0^{\infty}dp \, R_{n_bL_b}(p) R_{n_aL_a}(p) p^3$ by
$ \mu (m_a - m_b) \int _0^{\infty}dr\, R_{n_bL_b}(r) R_{n_aL_a}(r) r^3 $.
But, in so doing, one must be sure of two things : i) the quark-antiquark potential
does not depend on velocity (this is the case in our calculations) ii) the meson
wave functions are the true eigenstates of $H_{int}$. Even in this case
$(m_a - m_b)$ is the calculated value with a given potential and not the
experimental value.

Moreover, we find also a very popular additional approximation which consists in the
replacement of $E_0 (m_a - m_b)^2$ appearing in the width by $E_0^3$. This is
acceptable in a non relativistic situation but not in a relativistic one. All these
approximations are of course the remnants of the theory applied in atomic or
nuclear physics. We also discuss this point in the next section.
 
In the rest of the text this approximation will be called the dipole 
approximation (DA).

\subsection{general case}
We now come to the expression of the width in the general case, but with the wave
functions expanded on gaussian terms, as discussed in details previously. We present
the results only in the case $L_b=0$. The results for $L_a=0$ are easily obtained
from these ones with the correct replacement (\ref{emu2}) and the modification due 
to spin average. We do not want to enter
into too much details because the calculations are essentially a tricky application
of Racah algebra. We report the result below :

\begin{equation}
\label{larg}
\Gamma_{A \rightarrow B \gamma}=\delta_{S_b,J_b} 2 \alpha \frac{E_B(E_0)}{m_a}
[\mathcal{EE}+\mathcal{EM}+\mathcal{MM}]
\end{equation}

The terms $\mathcal{EE},\mathcal{EM},\mathcal{MM}$ come from the electric-electric, electric-
magnetic, magnetic-magnetic part in the square of the amplitude. One sees that a 
transition is no longer of purely electric or magnetic type, but a mixing of both
with interference effects. They are given by :
\begin{eqnarray}
\label{calem}
 \mathcal{EE}&=&\delta_{S_a,S_b} \frac{\Gamma(5/2)^2}{36}L_a(L_a+1)k_0^{2L_a-1}\mathcal{A}^2
\nonumber \\
 \mathcal{EM}&=&\delta_{S_a,S_b} (-1)^{L_a+S_a+J_a+J_b+1} \frac{\Gamma(3/2) \Gamma(5/2)}
{2 \sqrt{6}} \sqrt{L_a(L_a+1)} \hat{S_a} \hat{L_a} \hat{J_b} \nonumber \\
&& \left\{ \begin{tabular}{ccc} 
1 & 1/2 & 1/2 \\
1/2 & $S_a$ & $J_{b}$\\
\end{tabular}
\right\} 
\left\{ \begin{tabular}{ccc} 
$L_{a}$ & $S_{a}$ & $J_{a}$ \\
$J_{b}$ & $L_a$ & 1\\
\end{tabular} \right\}
k_0^{2L_a+1}\mathcal{A}\mathcal{B}  \\
 \mathcal{MM}&=&\frac{3}{4} \Gamma(3/2)^2 \hat{S_a}^2 \hat{L_a}^2 \hat{J_b}^2 
\left\{ \begin{tabular}{ccc} 
1 & 1/2 & 1/2 \\
1/2 & $S_a$ & $J_{b}$\\
\end{tabular}
\right\}^2
k_0^{2L_a+3}\mathcal{B}^2 \nonumber \\
&&\sum_{J} <1\; 1\; L_a\; 0 \mid J\; 1 >^2 
\left\{ \begin{tabular}{ccc} 
$L_a$ & $S_a$ & $J_a$ \\
$J_b$ & $J$ & 1\\
\end{tabular}
\right\}^2
\end{eqnarray}
where the dynamical factors $\mathcal{A}$ and $\mathcal{B}$ result from the meson
wave functions :
\begin{eqnarray}
 \mathcal{A}&=&\frac{< e_1 >}{m_1} \sum_{i,j} \frac{c_i^A c_j^B (x_{ij}^{(1)})^{L_a-1} 
\exp(-F_{ij}^{(1)}k_0^2)}{D_{ij}^{5/2}} \nonumber \\
&+&(-1)^{L_a} \frac{< e_2 >}{m_2} \sum_{i,j} \frac{c_i^A c_j^B (x_{ij}^{(2)})^{L_a-1} 
\exp(-F_{ij}^{(2)}k_0^2)}{D_{ij}^{5/2}} \\
 \mathcal{B}&=&\frac{< e_1 >}{m_1} \sum_{i,j} \frac{c_i^A c_j^B (x_{ij}^{(1)})^{L_a} 
\exp(-F_{ij}^{(1)}k_0^2)}{D_{ij}^{3/2}} \nonumber \\
&+&(-1)^{L_a+S_a+J_b} \frac{< e_2 >}{m_2} \sum_{i,j} \frac{c_i^A c_j^B 
(x_{ij}^{(2)})^{L_a} \exp(-F_{ij}^{(2)}k_0^2)}{D_{ij}^{3/2}}
\end{eqnarray}
Let us remark that if $L_a=0$ (transition from S state to S state), the terms
$\mathcal{EE}$ and $\mathcal{EM}$ vanishes, and the transition is purely magnetic.
Now, we have in hand all the tools to perform exact calculations. The formulae
look complicated but they simplify a lot for transitions of experimental
interest. Such simplified forms are presented in the appendix.

\section{The results}
We want to stress at the very beginning that, due to the fact that the wave functions
have been determined with given potentials and that the transition operator is
perfectly well defined, all the results presented in this section (except the very
last subsection concerning mixing angles) are free from any adjustable parameter.
They are thus a very good tool for exploring in detail the drawbacks of the formalism.

\subsection{\label{comments}Some comments on the mesons}
In the various potentials under consideration, we do not consider instanton
contribution which can mix strange and ordinary sectors.
So, in order to describe correctly the scalar $\eta$ mesons, we must introduce 
by hand some flavor mixing angle. Here we consider the idealized case of
equal mixing between $n$ flavor (isospin doublet $u$ or $d$) and strange flavor $s$,
which gives: 
$$\eta=\frac{(n\bar{n})_{I=0} -s\bar{s}}{\sqrt{2}}$$
$$\eta^{'}=\frac{(n\bar{n})_{I=0} +s\bar{s}}{\sqrt{2}}$$

\noindent With those presciptions the  $\eta$ and $\eta^{'}$ mesons are orthogonal.

Some other comments are in order. The QED conserves the flavor of the particles
at the vertex (the radiative transitions with flavor change $b \rightarrow s \gamma$
are intensively studied recently see \cite{Ell93,Bar98} but they need penguin
diagrams that we do not consider here); this implies that the quark content of
meson $B$ is the same as the one in the initial meson $A$ (in our elementary decay
process exhibited in figure (\ref{schema})).
Nevertheless the experimental data show a non zero decay width for the following
transitions concerning vector mesons: $\phi \rightarrow \omega \gamma$ and 
$\phi \rightarrow \rho \gamma$. 
For a $\phi$ meson taken to be a pure $s\bar{s}$, as usually prescribed, this 
seems to indicate that flavor conservation is violated. Instanton effects cannot
be advocated because they do not play any role for a spin triplet.
Those reactions result from more complicated processes. One can imagine for
instance an annihilation of the $q\bar{q}$ pair in meson $A$ into one or several
virtual gluons and creation of a new $q\bar{q}$ pair in meson $B$ (of possibly
different flavor) that interacts and emits a real photon. This is possible
only for neutral flavor mesons. Indeed such transitions
(with change of flavor) does not occur in kaons for example.

 A possible way to take into account phenomenologically those kinds 
of transitons, in our elementary process, is to include in the wave functions of
the $\omega$ a strange flavor part, or/and in the $\phi$ wave function 
a $n$ flavor part. A difficulty immediatly appears in the case of $\rho$ 
resonance since it is isovector while a strange flavor can only create an isoscalar.
For the $\omega$ and $\phi$ one could introduce a mixing flavor angle $\theta_f$. The
physical mesons are now a combination of the ideal $\phi_0 = s\bar{s}$ 
and $\omega_0=(n\bar{n})_{I=0}$.
$$\begin{array}{cccc}\label{phi_om}
\left(\begin{array}{c} 
\omega \\ \phi \\
\end{array}  \right) & = &
 \left(\begin{array}{cc} 
\cos \theta_f & \sin \theta_f \\
-\sin \theta_f & \cos \theta_f \\
\end{array}  \right) &
\left(\begin{array}{c} 
\omega_0 \\ \phi_0 \\
\end{array} \right) \\
\end{array}  $$

Now with this modification, the transition $\phi \rightarrow \omega \gamma$ can be
understood as the result of two contributions : 
$\phi(s\bar{s}) \rightarrow \omega(s\bar{s}) \gamma$ and
$\phi(n\bar{n}) \rightarrow \omega(n\bar{n}) \gamma$.
For the $\phi \rightarrow \rho \gamma$ only the $n$ flavor part of the $\phi$
contributes. 

From the electromagnetic point of view, only the charge is conserved, that is the
projection of the isospin but not the isospin value. We can imagine an isospin mixing
angle $\theta_I$ between neutral $\rho$ and $\omega$. As a consequence $\rho$ couples
to $\phi$ at the second order. Another evidence for this possible mixing angle is the
discrepancy between the neutral and charged channel of the
$\rho \rightarrow \pi \gamma$. This isospin mixing could be understood by the near
 degeneracy of $\rho$  and $\omega$ masses \cite{Pan97}.

$$\begin{array}{cccc}
\left(\begin{array}{c} 
\rho \\ \omega \\
\end{array}  \right) & = &
 \left(\begin{array}{cc} 
\cos \theta_I & \sin \theta_I \\
-\sin \theta_I & \cos \theta_I \\
\end{array}  \right) &
\left(\begin{array}{c} 
\rho_0(I=1) \\ \omega_0(I=0) \\
\end{array} \right) \\
\end{array}  $$

Our mixing angles are chosen in order to recover the usual prescription for
$\theta=0$; this does not correspond always to what can be found elsewhere. 
For a great part of our study, we work in the usual scheme and do not consider
mixing angles. We will drop those presciptions at the very end.

\subsection{Influence of the quality for the wave function}

We investigate the quality of our results as the number of
gaussians terms $N$ in the expansion (\ref{fop}) of the wave function varies
from $N=1$ (approximation) to $N=5$ (exact wave function). For this study, we use a
relativistic phase space, the AL1 potential and focus on the transition 
$ \rho^+ \rightarrow \pi^+ \gamma $. The convergence properties on the energies have
already been discussed in Table (\ref{mpiro}) . The two formalisms -- LWLA
and the general case -- are employed. The results are shown in the table (\ref{nbg}).

\begin{table}[htb]\begin{center}
\caption{\label{nbg}Variation of the decay width (in keV) as function of the number of gaussian
 terms used in the description of the meson wave functions. Long Wave Length
 Approximation (LWLA) and general case (Beyond LWLA), relativistic phase space and AL1
 potential are reported.}
\vspace{0.5cm}
\small{\begin{tabular}{|c||c|c|c|c|c||c|}
\hline

$\rho^+ \rightarrow \pi^+ \gamma$ & 1G & 2G & 3G & 4G & 5G  & exp.\\ \hline
 LWLA	& 50.98  & 56.57 & 56.98 & 56.98  & 57.01  & 67.82 $ \pm $ 7.55 \\ \hline
Beyond LWLA & 44.54  & 48.25 & 48.50 & 48.47  & 48.51  & 67.82 $ \pm $ 7.55 \\ \hline

\end{tabular}}\end{center}
\end{table}

First of all, one can see a sizeable improvement going from $N=1$ to $N=2$
for both studies. Limiting oneself to $N=1$, as is very often done in literature
can lead to 10 \% error. For $N=2$ to $N=3$ there is still a further small variation.
For $N=3$  the convergence is achieved and a stable result is reached; no
further improvement is got passing from $N=3$ to $N=5$.
This is consistent with figure (\ref{fig1}).
This means that 3 gaussian terms are enough to describe the wave function 
in a correct way. Nevertheless, for the rest of our study, every calculation 
is performed with $N=5$ and the result is considered as the exact one 
concerning the wave function. This is particularly important in the
case of radial excitations (with one or several nodes in the radial part), since 
obviously $N=1$ cannot explain such a state and even $N=2$ could be a crude
approximation. Moreover, since our results are analytical, computation with $N=5$
is practically as fast as the $N=3$ case.

We would like to point out that even if we used only one transition
($^3S_1 \rightarrow \: ^1S_0$) to test the quality of the wave function, the same conclusions
are valid whatever the transition under consideration. We know by experience that $N=5$
is always a very convenient tool to describe the exact wave function. The fact that
the above transition is better reproduced in LWLA than in the exact formalism is not
general and is commented later on.

\subsection{\label{phase} The phase space}

To study which phase space, among the three ones proposed in the last section, is the
most appropriate for those calculations, we used the AL1 potential, exact wave
functions and LWLA. In any case, we showed in other works that all phase space
considerations must be done with the experimental values of the resonance energies.
The transitions $^3P_J \rightarrow \: ^1S_0$ and $^1P_1 \rightarrow \: ^3S_1$ that
are seen experimentally cannot been explained in the LWLA and are not reported
in this part.

The results for the relativistic phase space (RPS), the mixed one (MPS) and the non 
relativistic one (NRPS) are compiled in the Tables (\ref{glo_rel_AL1}-\ref{glo_rel_AL1_2}).
A global glance at the quality of the results vs the phase space immediatly
reveals that (NRPS) is not a correct prescription, the theoretical results being
very often too large, except may be for heavy mesons for which a non relativistic
treatment is acceptable. The problem comes essentially from a wrong determination
of the momentum $k_0=E_0$, that enters both in the $\frac{E_B}{m_A}$ factor
(here taken as unity) and in the dynamical amplitudes. There exist some important
discrepancies between the relativistic and the non relativistic momenta
(this could be seen in the second and third columns of the different tables). We remind you
that the experimental values of the meson masses are used. The MPS supports
this conclusion; it uses the relativistic impulsion but a $\frac{E_B}{m_A}$ factor
equal to one (as in the non relativistic phase space) and the results are much 
better than NRPS ones, and little poorer than RPS ones. So the predicted decays
widths in NRPS are essentially spoiled by wrong values of momentum.
The change in the results going from MPS to RPS is proportional to the factor
$\frac{E_B}{m_A}$. So looking at the fourth column of the tables, one sees that
the biggest deviation is around a factor 2 and concerns the transitions :
$\rho^+ \rightarrow \pi^+ \gamma$ or $\omega\rightarrow \pi^0\gamma$. 
The use of RPS is quite satisfactory for the whole bulk of data, indicating that
the wave functions are not completely crazy.

In the fourth column we also see that $\frac{E_B}{m_A}=1$ is a justified approximation 
for the heavy mesons, this is due to the fact that the mass spectrum is here more dense 
than in the light sector.

With those values of momenta and considering their important values for 
some transitions, it is also interesting  to be aware that the LWLA
could not be always a justified formalism. In the case of important mass
differences between initial and final state, the recoil term could not be omitted.

In view of this discussion, the rest of our study will be done using a relativistic
phase space.

\begin{table}[htb]\begin{center}
\caption{\label{glo_rel_AL1} Decay widths obtained through the Long Wave Length Approximation,
relativistic (rel), mixed and non relativistic (nrel) phase spaces, AL1 potential. The momentum 
$k_0=E_0$ is expressed in MeV whereas the decay widths are in keV. The mixed phase space uses a
relativistic momentum contrary to the non relativistic one which also uses a ratio $\frac{E_B}{m_A}$
equal to the unity.}
\begin{tabular}{|l||c|c|c||c|c|c||c|}
\hline
\multicolumn{8}{|c|}{ } \\  
\multicolumn{8}{|c|}{$^3S_1 \rightarrow \: ^1S_0$} \\  
\multicolumn{8}{|c|}{ } \\  \hline
transition &  $k_0^{rel}$ & $k_0^{nrel}$ & $\frac{E_B}{m_A}$ & rel.  & mixed  & nrel.  & exp. \\ \hline
$ \rho^+ \rightarrow \pi^+ \gamma $        & 372 & 630 & 0.52 & 56.98  & 110.32 & 535.23 & 67.82 $ \pm $ 7.55 \\ \hline
$ \rho^0 \rightarrow \pi^0 \gamma $        & 373 & 634 & 0.52 & 57.24  & 111.05 & 547.02 & 102.48$ \pm $ 25.69 \\ \hline
$ \rho^0 \rightarrow \eta \gamma  $        & 190 & 222 & 0.75 & 49.79  & 66.12 & 105.52 & 36.18 $ \pm $ 13.57 \\ \hline
$ \omega \rightarrow \pi^0 \gamma $        & 380 & 648 & 0.51 & 543.44 & 1055.48 & 5238.71& 714.85 $ \pm $ 42.74 \\ \hline
$ \omega \rightarrow \eta \gamma  $        & 200 & 235 & 0.74 & 6.37   & 8.56 & 13.96  & 5.47 $ \pm $ 0.84 \\ \hline
$ \phi(1020) \rightarrow \eta \gamma   $   & 363 & 472 & 0.64 & 45.90  & 71.26 & 157.04 & 55.82 $ \pm $ 2.73 \\ \hline
$ \phi(1020) \rightarrow \eta'(958)\gamma$ & 60  & 62  & 0.94 & 0.30   & 0.32 & 0.35   & 0.53 $ \pm $ 0.31 \\ \hline
$ K^*(892)^0 \rightarrow K^0 \gamma    $   & 310 & 398 & 0.65 & 111.19 & 169.95 & 361.34 & 116.15 $ \pm $ 10.19 \\ \hline
$ K^*(892)^+ \rightarrow K^+ \gamma    $   & 309 & 398 & 0.65 & 97.47  & 149.21 & 318.28 & 50.29 $ \pm $ 4.66 \\ \hline
$ D^*(2007)^0 \rightarrow D^0 \gamma   $   & 137 & 142 & 0.93 & 35.08  & 37.65 & 41.95 &$<$ 800.10 $ \pm $ 60.90 \\ \hline
$ D^*(2010)^+ \rightarrow D^+ \gamma   $   & 136 & 141 & 0.93 & 2.68   & 2.87 & 3.19  &$<$ 1.44 $^{+2.75}_{-0.92}$\\ \hline
$ D_s^{*+} \rightarrow D_s^+ \gamma    $   & 139 & 144 & 0.93 & 0.28   & 0.29 & 0.33  & $<$ 1789.80 $ \pm $ 47.50 \\ \hline
$ B^{*+} \rightarrow B^+ \gamma  $         & 46  & 46  & 0.99 & 0.97   & 0.98 & 0.99  & seen  \\ \hline
$ B^{*0} \rightarrow B^0 \gamma  $         & 45  & 46  & 0.99 & 0.28   & 0.29 & 0.29  & seen   \\ \hline
$ J/\Psi  \rightarrow \eta_c  \gamma  $    & 115 & 117 & 0.96 & 1.85   & 1.93 & 2.04  & 1.13 $ \pm $ 0.35 \\ \hline
$ \psi(2S) \rightarrow \eta_c(1S) \gamma$  & 639 & 706 & 0.83 & 2.15   & 2.60 & 3.52  & 0.78 $ \pm $ 0.19\\ \hline \hline

\multicolumn{8}{|c|}{ } \\  
\multicolumn{8}{|c|}{$^1S_0 \rightarrow \: ^3S_1$ } \\  
\multicolumn{8}{|c|}{ } \\  \hline
transition &  $k_0^{rel}$ & $k_0^{nrel}$ & $\frac{E_B}{m_A}$ & rel.  & mixed  & nrel.  & exp. \\ \hline
$ \eta'(958) \rightarrow \rho^0 \gamma $ & 170 & 188 & 0.82 & 116.80 & 141.99 & 193.73 & 61.31 $ \pm $ 5.51 \\ \hline
$ \eta'(958) \rightarrow \omega \gamma $ & 159 & 175 & 0.83 & 10.81  & 12.97 & 17.29 & 6.11 $ \pm $ 0.78 \\ \hline\hline

\end{tabular}\end{center}
\end{table}

\begin{table}[htb]\begin{center}
\caption{\label{glo_rel_AL1_2}Same as Table (\ref{glo_rel_AL1}).}
\begin{tabular}{|l||c|c|c||c|c|c||c|}
\hline
\multicolumn{8}{|c|}{ } \\  
\multicolumn{8}{|c|}{$^3S_1 \rightarrow \: ^3P_J$ } \\  
\multicolumn{8}{|c|}{ } \\  \hline
transition &  $k_0^{rel}$ & $k_0^{nrel}$ & $\frac{E_B}{m_A}$ & rel.  & mixed  & nrel.  & exp. \\ \hline
$ \psi(2S) \rightarrow \chi_{c0}(1P) \gamma $ & 261 & 271 & 0.93 & 17.70 & 19.04 & 19.77 & 25.76$ \pm $ 3.81\\ \hline
$ \psi(2S) \rightarrow \chi_{c1}(1P) \gamma $ & 171 & 175 & 0.95 & 35.75 & 37.49 & 38.40 & 24.10 $ \pm $ 3.49\\ \hline
$ \psi(2S) \rightarrow \chi_{c2}(1P) \gamma $ & 127 & 130 & 0.97 & 44.90 & 46.51 & 47.35 & 21.61 $ \pm $ 3.28 \\ \hline
$ \Upsilon(2S) \rightarrow \chi_{b0}(1P) \gamma $ & 162 & 163 & 0.98 & 0.43 & 0.44 & 0.44 & 1.89 $ \pm $ 0.53 \\ \hline
$ \Upsilon(2S) \rightarrow \chi_{b1}(1P) \gamma $ & 130 & 131 & 0.99 & 1.04 & 1.05 & 1.06 & 2.95 $ \pm $ 0.61 \\ \hline
$ \Upsilon(2S) \rightarrow \chi_{b2}(1P) \gamma $ & 110 & 111 & 0.99 & 1.47 & 1.48 & 1.49 & 2.90 $ \pm $ 0.61 \\ \hline
$ \Upsilon(3S) \rightarrow \chi_{b0}(2P) \gamma $ & 122 & 123 & 0.99 & 0.68 & 0.69 & 0.70 & 1.42 $ \pm $ 0.25 \\ \hline
$ \Upsilon(3S) \rightarrow \chi_{b1}(2P) \gamma $ & 100 & 100 & 0.99 & 1.67 & 1.69 & 1.69 & 2.97 $ \pm $ 0.43 \\ \hline
$ \Upsilon(3S) \rightarrow \chi_{b2}(2P) \gamma $ & 86  & 87  & 0.99 & 2.42 & 2.44 & 2.45 & 3.00 $ \pm $ 0.45\\ \hline\hline
\multicolumn{8}{|c|}{ } \\  
\multicolumn{8}{|c|}{$^3P_J \rightarrow \: ^3S_1$ } \\  
\multicolumn{8}{|c|}{ } \\  \hline
transition &  $k_0^{rel}$ & $k_0^{nrel}$ & $\frac{E_B}{m_A}$ & rel.  & mixed  & nrel.  & exp. \\ \hline
$ f_1(1285) \rightarrow \rho^0 \gamma  $  	  & 410 & 513 & 0.68 & 939.73 & 1381.81 & 1727.12  & 1296.00 $ \pm $ 295.20  \\ \hline 
$ \chi_{c0}(1P) \rightarrow J/\psi(1S) \gamma $   & 303 & 318 & 0.91 & 233.40 & 256.15 & 268.67 & 92.40 $ \pm $ 41.52 \\ \hline
$ \chi_{c1}(1P) \rightarrow J/\psi(1S) \gamma $   & 389 & 414 & 0.89 & 292.29 & 328.75 & 349.33 & 240.24 $ \pm $ 40.73 \\ \hline
$ \chi_{c2}(1P) \rightarrow J/\psi(1S) \gamma $   & 430 & 459 & 0.88 & 319.01 & 362.85 & 387.90 & 270.00$\pm$32.78\\ \hline
$ \chi_{b0}(1P) \rightarrow \Upsilon(1S) \gamma $ & 392 & 400 & 0.96 & 28.82 & 30.01 & 30.63 & seen  \\ \hline
$ \chi_{b1}(1P) \rightarrow \Upsilon(1S) \gamma $ & 423 & 432 & 0.96 & 31.03 & 32.42 & 33.14 & seen  \\ \hline
$ \chi_{b2}(1P) \rightarrow \Upsilon(1S) \gamma $ & 442 & 452 & 0.96 & 32.37 & 33.88 & 34.67 & seen  \\ \hline
$ \chi_{b0}(2P) \rightarrow \Upsilon(1S) \gamma $ & 743 & 772 & 0.93 & 12.12 & 13.07 & 13.58 & seen  \\ \hline
$ \chi_{b0}(2P) \rightarrow \Upsilon(2S) \gamma $ & 207 & 209 & 0.98 & 13.07 & 13.34 & 13.47 & seen  \\ \hline
$ \chi_{b1}(2P) \rightarrow \Upsilon(1S) \gamma $ & 764 & 795 & 0.93 & 12.45 & 13.45 & 13.99 & seen  \\ \hline
$ \chi_{b1}(2P) \rightarrow \Upsilon(2S) \gamma $ & 229 & 232 & 0.98 & 14.47 & 14.80 & 14.97 & seen  \\ \hline
$ \chi_{b2}(2P) \rightarrow \Upsilon(1S) \gamma $ & 776 & 808 & 0.92 & 12.63 & 13.66 & 14.22 & seen  \\ \hline
$ \chi_{b2}(2P) \rightarrow \Upsilon(2S) \gamma $ & 242 & 245 & 0.98 & 15.27 & 15.63 & 15.82 & seen  \\ \hline

\multicolumn{8}{|c|}{ } \\  
\multicolumn{8}{|c|}{$^1P_1 \rightarrow \:^1S_0$ } \\  
\multicolumn{8}{|c|}{ } \\  \hline
transition &  $k_0^{rel}$ & $k_0^{nrel}$ & $\frac{E_B}{m_A}$ & rel.  & mixed  & nrel.  & exp. \\ \hline
$ b_1(1235)^+ \rightarrow \pi^+ \gamma $ & 607 & 1090 & 0.51 & 209.95  & 414.55 & 744.58 & 227.20 $ \pm $ 58.60 \\ \hline\hline
\end{tabular}\end{center}
\end{table}

\subsection{Dipole approximation}

In this part we want to discuss the dipole approximation (DA) for electric transition.
A relativistic phase space is used, but we study two different prescriptions suggested
previously. DA1 is the expression resulting from the formalism : it appears a term
$E_0 (m_a - m_b)^2$ in phase space; DA2 is obtained with this term replaced by $E_0^3$.
The results are presented in the table (\ref{sieg1}). It is difficult to draw some 
conclusions , mainly because of lack of precise experimental data.
We first remark a small difference (less than 15\%) between the DA1 and DA2. 
This is due to the fact that almost all exploitable data concern heavy mesons for
which the two approximations tend to be the same. For example, the transition  
$\chi_{b2}(2P)\rightarrow \Upsilon(1S)\gamma$  gives a momentum value of $E_0=776$ MeV 
for $m_a - m_b=808$ MeV, that is a  8\% variation. Indeed only two transitions deal
with ordinary quarks : $ f_1(1285) \rightarrow \rho^0 \gamma $ and 
$b_1(1235)^+ \rightarrow \pi^+ \gamma$ and in this case the difference is appreciable.

Generally speaking the DA gives better agreement than LWLA on weak transitions 
(around 1 keV) but this is not significant. On the contrary the results are worse
in the case of strong transitions. Although the DA is very popular in atomic and nuclear
physics -- a domain where it can be applied safely -- it should be avoided in the
meson sector.

\begin{table}[htb]\begin{center}
\caption{\label{sieg1} Dipole approximation in the LWLA, AL1 potential and
relativistic phase space. Two approximations are presented depending upon
the replacement (DA2) or not (DA1) of $(m_B-m_A)$ by $E_0$, see text.}
\small{\begin{tabular}{|l||c|c||c|c||c|}
\hline
% \multicolumn{6}{|c|}{ } \\  
\multicolumn{6}{|c|}{$ ^3S_1 \rightarrow \: ^3P_J$ } \\  
\multicolumn{6}{|c|}{ } \\  \hline
transition & $k_0$[MeV] & $\frac{E_B}{m_A}$ &  DA1 & DA2. &  exp.[keV]\\ \hline
$ \psi(2S) \rightarrow \chi_{c0}(1P) \gamma   $   & 261 & 0.93 & 49.92 & 53.80 & 25.76$ \pm $ 3.81\\ \hline
$ \psi(2S) \rightarrow \chi_{c1}(1P) \gamma   $   & 171 & 0.95 & 43.43 & 45.57 & 24.10$ \pm $ 3.49\\ \hline
$ \psi(2S) \rightarrow \chi_{c2}(1P) \gamma   $   & 127 & 0.97 & 30.23 & 31.32 & 21.61$ \pm $ 3.28 \\ \hline
$ \Upsilon(2S) \rightarrow \chi_{b0}(1P) \gamma $ & 162 & 0.98 & 1.52  & 1.54  & 1.89 $ \pm $ 0.53 \\ \hline
$ \Upsilon(2S) \rightarrow \chi_{b1}(1P) \gamma $ & 130 & 0.99 & 2.34  & 2.37  & 2.95 $ \pm $ 0.61 \\ \hline
$ \Upsilon(2S) \rightarrow \chi_{b2}(1P) \gamma $ & 110 & 0.99 & 2.39  & 2.41  & 2.90 $ \pm $ 0.61 \\ \hline
$ \Upsilon(3S) \rightarrow \chi_{b0}(2P) \gamma $ & 122 & 0.99 & 1.65  & 1.67  & 1.42 $ \pm $ 0.25 \\ \hline
$ \Upsilon(3S) \rightarrow \chi_{b1}(2P) \gamma $ & 100 & 0.99 & 2.67  & 2.69  & 2.97 $ \pm $ 0.43 \\ \hline
$ \Upsilon(3S) \rightarrow \chi_{b2}(2P) \gamma $ & 86  & 0.99 & 2.91  & 2.93  & 3.00 $ \pm $ 0.45 \\ \hline
% \multicolumn{6}{|c|}{ } \\  
\multicolumn{6}{|c|}{$ ^3P_J \rightarrow \: ^3S_1$ } \\  
\multicolumn{6}{|c|}{ } \\  \hline
transition & $k_0$ & $\frac{E_B}{m_A}$ &  DA1 & DA2 &  exp.\\ \hline
$ f_1(1285) \rightarrow \rho^0 \gamma     $       & 410 & 0.68 & 824.56 & 1288.17& 1296.00$ \pm $ 295.20\\ \hline 
$ \chi_{c0}(1P) \rightarrow J/\psi(1S) \gamma $   & 303 & 0.91 & 144.59 & 159.06 & 92.40  $ \pm $ 41.52 \\ \hline
$ \chi_{c1}(1P) \rightarrow J/\psi(1S) \gamma $   & 389 & 0.89 & 298.24 & 336.75 & 240.24 $ \pm $ 40.73 \\ \hline
$ \chi_{c2}(1P) \rightarrow J/\psi(1S) \gamma $   & 430 & 0.88 & 396.53 & 453.17 & 270.00 $ \pm $ 32.78 \\ \hline
$ \chi_{b0}(1P) \rightarrow \Upsilon(1S) \gamma $ & 392 & 0.96 & 20.36  & 21.21  & seen  \\ \hline
$ \chi_{b1}(1P) \rightarrow \Upsilon(1S) \gamma $ & 423 & 0.96 & 25.59  & 26.74  & seen  \\ \hline
$ \chi_{b2}(1P) \rightarrow \Upsilon(1S) \gamma $ & 442 & 0.96 & 29.14  & 30.52  & seen  \\ \hline
$ \chi_{b0}(2P) \rightarrow \Upsilon(1S) \gamma $ & 743 & 0.93 & 11.32  & 12.22  & seen  \\ \hline
$ \chi_{b0}(2P) \rightarrow \Upsilon(2S) \gamma $ & 207 & 0.98 & 9.37   & 9.56   & seen  \\ \hline
$ \chi_{b1}(2P) \rightarrow \Upsilon(1S) \gamma $ & 764 & 0.93 & 12.30  & 13.31  & seen  \\ \hline
$ \chi_{b1}(2P) \rightarrow \Upsilon(2S) \gamma $ & 229 & 0.98 & 12.77  & 13.06  & seen  \\ \hline
$ \chi_{b2}(2P) \rightarrow \Upsilon(1S) \gamma $ & 776 & 0.92 & 12.89  & 13.97  & seen  \\ \hline
$ \chi_{b2}(2P) \rightarrow \Upsilon(2S) \gamma $ & 242 & 0.98 & 15.04  & 15.41  & seen  \\ \hline
% \multicolumn{6}{|c|}{ } \\  
\multicolumn{6}{|c|}{$ ^1P_1 \rightarrow \: ^1S_0$ } \\  
\multicolumn{6}{|c|}{ } \\  \hline
transition & $k_0$ & $\frac{E_B}{m_A}$ &  DA1 & DA2 &  exp.\\ \hline
$b_1(1235)^+ \rightarrow \pi^+ \gamma$ & 607 & 0.51 & 84.33 & 272.06 & 227.20 $\pm$ 58.60\\ \hline

\end{tabular}}\end{center}
\end{table}

\subsection{General study for three different quark-antiquark potentials }

In this part we present the most sophisticated calculations in this framework. We
go beyond LWLA, use relativistic phase space and exact wave functions. The objective
is twofold. First to test under which conditions a general treatment is necessary
as compared to the often used LWLA, and how big could be the difference. Second,
by testing the three quark-antiquark potentials proposed in section \ref{ssec:pot},
to see whether the results are very sensitive to the dynamics of quarks inside a meson.
The results are presented in tables (\ref{rel_AL1}-\ref{rel_AL1_2}-\ref{rel_AL1_3}).

Let us compare this general case with the LWLA. Several comments are in order.
In a number of cases there is no electric-magnetic mixing.
For $L_a=L_b=0$ only the magnetic term remains while a pure 
electric term remains for $^1P_1 \rightarrow \: ^1S_0$. In this case the LWLA gives
always a larger value; this can be shown theoretically if the wave function has only
one gaussian component. Since in general one component is dominant it is not
surprising that this property persists even in more realistic situations. Although
this is not always the case, LWLA often gives better (as compared to experiment)
results; this means that either the wave functions are not so good or that something is
still missing in the theory.

The transitions $^1P_1 \rightarrow \: ^3S_1$ and $^3P_J \rightarrow \: ^1S_0$ are also
purely magnetic but are completely forbidden in LWLA. Our complete treatment predicts
them with the right order of magnitude.

Lastly $^3P_J\rightarrow \: ^3S_1$ and $ ^3S_1\rightarrow \: ^3P_J$ transitions
show interference between electric and magnetic contributions while they are
purely electric in LWLA. Since the electric part remains largely dominant there
is no big difference between both formalisms except in the case of 
$f_1(1285) \rightarrow \: \rho \gamma$ which is greatly enhanced in the right
direction.

An overall look at those tables shows us the similarity of the decay widths 
resulting from the AL1, AP1 and Bhaduri potentials. The results obtained
with AL1 generally lie between those of Bhaduri and AP1. The predicted values 
coming from the AL1 potential are smaller than the AP1 ones. This could be 
related to the different asymptotic behavior of the potential at long range. 
The confinement being slower in the AP1 potential, the spread of the wave 
function is more important and contributes more in the spacial integration.
Globally no potential is really more suited than the other for those calculations,
although Badhuri's one seems a little poorer. The trends are essentially the
same and when one potential gives too low (or too high) value, so do the others.
The agreement with experiment is satisfactory for all types of transitions,
giving indication that we are in the good track for the description of mesons.
The discrepancy with experimental situation exceeds 50\% very scarcely and
this can be considered as encouraging but not completely satisfactory.

The fact that three different wave functions give more or less the same trends
(although there can exist 20\% differences) shows that the quality of the wave
function is not responsible entirely for the discrepancy with experiment.
Moreover LWLA as an approximation should give poorer agreement than a complete
exact treatment. This is not the case (except of course when it gives a null
result). This proves that something is still missing in the formalism although
the present day calculations provide the dominant contributions. 
\\

Now let us have a closer look to some interesting  transitions. A special one concerns
the neutral and charged decays of the $\rho$ into the $\pi \gamma$.  Experimentally 
the decay width for the charged channel is 68 keV whereas the measure for the neutral is 
102 keV. In our calculation the small difference between those two channels comes only 
from the tiny difference between the experimental masses of the $\pi^+$ and the $\pi^0$.
The decay width has the same expression for those transitions due to the term 
$ \left[ \frac{\langle e_1\rangle}{m_1} -  \frac{\langle e_2 \rangle}{m_2}\right]^2$
in expression (\ref{mmu}) appearing for meson composed of a single flavor.
That is $[\frac{2}{3}-\frac{1}{3}]^2=\frac{1}{9}$ for the charged channel and
$[\frac{1}{6}-\frac{-1}{6}]^2=\frac{1}{9}$ for the neutral one.
So where does this important variation between those two channels come from ?
First we have to point out that given the large uncertainties  67.82 $\pm$ 7.55 and 
 102.48 $\pm$ 25.69 the two values are nearly compatible with 76 MeV. This 
means that there may be no problem with those channels except an experimental one !
Nevertheless if we rely more deeply on the experimental values a possible
insight could come from the $\omega \rightarrow \pi^0 \gamma$ transition, 
which is identical to $\rho^0 \rightarrow \pi^0 \gamma$ but with an enhancement by
a factor 9 due to isospin. The experimental value 715 keV is roughly in agreement
with this point. So even a small isospin mixing 
between the $\omega$ and the $\rho^0$ could increase sufficiently the decay width 
to explain the data. This hypothesis will be tested in the next part.

We remark the same variation for the K* decaying into the K meson and for the B* into
the B. Yet this time the isospin factor explains this variation. It appears a factor 
$(\frac{1}{3m_n}+\frac{1}{3m_s})^2$ for the neutral channel and a factor 
$(\frac{2}{3m_n}-\frac{1}{3m_s})^2$ for the charged one in the LWLA (it is more 
complicated to estimate the ratio of the strange with the isospin doublet masses 
in the general formalism ). Using the experimental data and making the approximation that 
the matrix elements in those two channels are identical (except the isospin dependence),
we find the relation : 
$ m_s = 1.24 m_n$. In our potentials this ratio $\frac{m_s}{m_n}$ is 1.83, 1.78, 2.00 
for the AL1, AP1 and Bhaduri respectively.\\

Concerning the transition $a_1^0\rightarrow \pi^0\gamma $, the decay width is zero; this is
due to the fact that for the $^3P_J \rightarrow \: ^1S_0$ composed of a single flavor 
the width is propotional to $ \left[ \frac{\langle e_1\rangle}{m_1} +
 \frac{\langle e_2 \rangle}{m_2}\right]=0$.\\
In the potentials used, there is no isospin dependence so the $\rho, \omega$ have the 
same radial part of the wave function, and the same remark is true for the $\pi$ and
$\eta_n$ (no instanton effect). This means that for the transitions number one to five
of the $^3S_1 \rightarrow \: ^1S_0$ tables, the integral part is identical. Now since
 the experimental hierarchy of the decay widths is reproduced in our calculation this
could indicate that we used a correct prescription; moreover since the numerical value are 
also reproduced this ensures the quality of our wave functions.\\
It is not sure that the decay of the $D_{s1}(2536)^{*+}$ into $D_{s}^{*}\gamma$
has been observed experimentaly but our result for this width tends to prove that
 it should be seen experimentally.\\

\begin{table}[htb]\begin{center}
\caption{\label{rel_AL1} General case, relativistic phase space, AL1, AP1 and Bhaduri's potentials.
For the values of $k_0^{rel}$ and $\frac{E_B}{m_A}$ see Tables (\ref{glo_rel_AL1}-\ref{glo_rel_AL1_2}).
The electric, magnetic and interference terms refer to the AL1 potential.}

\small{\begin{tabular}{|l||c|c|c||c||c||c||c|}
\hline
\multicolumn{8}{|c|}{ } \\
\multicolumn{8}{|c|}{ $ ^3S_1\rightarrow \: ^1S_0 $} \\
\multicolumn{8}{|c|}{ } \\ \hline
transition &  elec. & interfer. & magn. & tot(AL1)& tot(AP1) & tot(BD) & exp. \\ \hline 
$ \rho^+ \rightarrow \pi^+ \gamma $        &  &  & 48.48 & 48.48 & 60.41 & 44.07 & 67.82 $ \pm $ 7.55 \\ \hline
$ \rho^0 \rightarrow \pi^0 \gamma $        &  &  & 48.66 & 48.66 & 60.64 & 44.24 & 102.48$ \pm $ 25.69 \\ \hline
$ \rho^0 \rightarrow \eta \gamma  $        &  &  & 47.73 & 47.73 & 60.63 & 42.78 & 36.18 $ \pm $ 13.57 \\ \hline
$ \omega \rightarrow \pi^0 \gamma $        &  &  & 459.30& 459.30& 571.79& 417.85 & 714.85 $ \pm $ 42.74 \\ \hline
$ \omega \rightarrow \eta \gamma  $        &  &  & 6.08  & 6.08  & 7.72  & 5.43 & 5.47 $ \pm $ 0.84 \\ \hline
$ \phi(1020) \rightarrow \eta \gamma   $   &  &  & 41.27 & 41.27 & 44.12 & 39.74 & 55.82 $ \pm $ 2.73 \\ \hline
$ \phi(1020) \rightarrow \eta'(958)\gamma$ &  &  & 0.30  & 0.30  & 0.32  & 0.29 & 0.53 $ \pm $ 0.31 \\ \hline
$ K^*(892)^0 \rightarrow K^0 \gamma    $   &  &  & 98.28 & 98.28 & 116.41& 91.70 & 116.15 $ \pm $ 10.19 \\ \hline
$ K^*(892)^+ \rightarrow K^+ \gamma    $   &  &  & 79.07 & 79.07 & 104.46& 71.50 & 50.29 $ \pm $ 4.66 \\ \hline
$ D^*(2007)^0 \rightarrow D^0 \gamma   $   &  &  & 33.60 & 33.60 & 41.74 & 30.24 & $<$ 800.10 $ \pm $ 60.90 \\ \hline
$ D^*(2010)^+ \rightarrow D^+ \gamma   $   &  &  & 2.48  & 2.48  & 3.58  & 2.09 & $<$ 1.44 $^{+2.75}_{-0.92}$ \\ \hline
$ D_s^{*+} \rightarrow D_s^+ \gamma    $   &  &  & 0.26  & 0.26  & 0.31  & 0.22 & $<$ 1789.80 $ \pm $ 47.50 \\ \hline
$ B^{*+} \rightarrow B^+ \gamma  $         &  &  & 0.97  & 0.97  & 1.26  & 0.84 & seen  \\ \hline
$ B^{*0} \rightarrow B^0 \gamma  $         &  &  & 0.28  & 0.28  & 0.36  & 0.25 & seen	\\ \hline
$ J/\Psi  \rightarrow \eta_c  \gamma  $    &  &  & 1.85  & 1.85  & 1.87  & 1.79 & 1.13 $ \pm $ 0.35 \\ \hline
$\psi(2S) \rightarrow \eta_c(1S) \gamma$   &  &  & 4.97  & 4.97  & 6.34  & 2.44 & 0.78 $ \pm $ 0.19 \\ \hline

\multicolumn{8}{|c|}{ }\\ 
\multicolumn{8}{|c|}{$^1S_0 \rightarrow \: ^3S_1$}\\ 
\multicolumn{8}{|c|}{ }\\ \hline
transition &  elec. & interfer. & magn. & tot(AL1)& tot(AP1) & tot(BD) & exp. \\ \hline 
$ \eta'(958) \rightarrow \rho^0 \gamma $   &  &  & 112.90& 112.90& 143.62& 101.09 & 61.31 $ \pm $ 5.51 \\ \hline
$ \eta'(958) \rightarrow \omega \gamma $   &  &  & 10.50 & 10.50 & 13.36 & 9.39 & 6.11 $ \pm $ 0.78 \\ \hline

\end{tabular}}\end{center}
\end{table}

\begin{table}[htb]\begin{center}
\caption{\label{rel_AL1_2}Same as Table (\ref{rel_AL1}).}

\small{\begin{tabular}{|l||c|c|c||c||c||c||c|}
\hline

\multicolumn{8}{|c|}{ }\\
\multicolumn{8}{|c|}{$ ^3S_1\rightarrow \: ^3P_J$}\\
\multicolumn{8}{|c|}{ }\\ \hline
transition &  elec. & interfer. & magn. & tot(AL1)& tot(AP1) & tot(BD) & exp. \\ \hline 
$ \psi(2S) \rightarrow \chi_{c0}(1P) \gamma $     & 18.00& -4.12 & 0.24 & 14.12& 14.06 & 14.94 & 25.76$ \pm $ 3.81\\ \hline
$ \psi(2S) \rightarrow \chi_{c1}(1P) \gamma $     & 36.02& -1.82 & 0.05 & 34.25& 34.23 & 35.70 & 24.10 $\pm $3.49\\ \hline
$ \psi(2S) \rightarrow \chi_{c2}(1P) \gamma $     & 45.09& 1.27  & 0.03 & 46.39& 46.43 & 48.07 & 21.61$\pm $3.28 \\ \hline
$ \Upsilon(2S) \rightarrow \chi_{b0}(1P) \gamma $ & 0.43 & -0.02 & 0.00 & 0.41 & 0.54  & 0.45 & 1.89 $ \pm $ 0.53 \\ \hline
$ \Upsilon(2S) \rightarrow \chi_{b1}(1P) \gamma $ & 1.04 & -0.02 & 0.00 & 1.02 & 1.35  & 1.13 & 2.95 $ \pm $ 0.61 \\ \hline
$ \Upsilon(2S) \rightarrow \chi_{b2}(1P) \gamma $ & 1.47 & 0.02  & 0.00 & 1.49 & 1.97  & 1.64 & 2.90 $ \pm $ 0.61 \\ \hline
$ \Upsilon(3S) \rightarrow \chi_{b0}(2P) \gamma $ & 0.68 & -0.02 & 0.00 & 0.66 & 0.73  & 0.72 & 1.42 $ \pm $ 0.25 \\ \hline
$ \Upsilon(3S) \rightarrow \chi_{b1}(2P) \gamma $ & 1.67 & -0.02 & 0.00 & 1.65 & 1.84  & 1.79 & 2.97 $ \pm $ 0.43 \\ \hline
$ \Upsilon(3S) \rightarrow \chi_{b2}(2P) \gamma $ & 2.42 & 0.02  & 0.00 & 2.44 & 2.71  & 2.64 & 3.00 $ \pm $ 0.45\\ \hline
\multicolumn{8}{|c|}{ }\\
\multicolumn{8}{|c|}{$^3P_J \rightarrow \: ^3S_1 $ }\\
\multicolumn{8}{|c|}{ }\\ \hline
transition &  elec. & interfer. & magn. & tot(AL1)& tot(AP1) & tot(BD) & exp. \\ \hline 
$ f_1(1285) \rightarrow \rho^0 \gamma $          & 688.97&417.42 &126.45&1232.83& 1376.96& 1209.32 & 1296.00 $\pm$295.20\\ \hline
$ \chi_{c0}(1P) \rightarrow J/\psi(1S) \gamma$   & 225.25& 29.20 & 0.95 & 255.40& 260.24 & 271.53 & 92.40 $\pm$41.52\\ \hline
$ \chi_{c1}(1P) \rightarrow J/\psi(1S) \gamma$   & 275.69& 29.38 & 1.57 & 306.63& 312.43 & 327.31 & 240.24 $\pm$40.73\\ \hline
$ \chi_{c2}(1P) \rightarrow J/\psi(1S) \gamma$   & 297.08& -38.52& 3.50 & 262.05& 266.99 & 286.79 & 270.00 $\pm$ 32.78\\ \hline
$ \chi_{b0}(1P) \rightarrow \Upsilon(1S) \gamma$ & 28.29 & 1.78  & 0.03 & 30.10 & 30.85 & 32.24 & seen\\ \hline
$ \chi_{b1}(1P) \rightarrow \Upsilon(1S) \gamma$ & 30.37 & 1.11  & 0.02 & 31.51 & 32.26 & 33.82 & seen\\ \hline
$ \chi_{b2}(1P) \rightarrow \Upsilon(1S) \gamma$ & 31.62 & -1.26 & 0.04 & 30.39 & 31.03 & 32.82 & seen\\ \hline
$ \chi_{b0}(2P) \rightarrow \Upsilon(1S) \gamma$ & 12.22 & 1.73  & 0.06 & 14.01 & 11.80 & 14.69 & seen\\ \hline
$ \chi_{b0}(2P) \rightarrow \Upsilon(2S) \gamma$ & 12.88 & 0.43  & 0.00 & 13.31 & 13.52 & 14.28 & seen\\ \hline
$ \chi_{b1}(2P) \rightarrow \Upsilon(1S) \gamma$ & 12.55 & 0.94  & 0.04 & 13.53 & 11.38 & 14.25 & seen\\ \hline
$ \chi_{b1}(2P) \rightarrow \Upsilon(2S) \gamma$ & 14.21 & 0.29  & 0.00 & 14.51 & 14.72 & 15.58 & seen\\ \hline
$ \chi_{b2}(2P) \rightarrow \Upsilon(1S) \gamma$ & 12.74 & -0.99 & 0.05 & 11.80 & 9.90  & 12.60 & seen\\ \hline
$ \chi_{b2}(2P) \rightarrow \Upsilon(2S) \gamma$ & 14.97 & -0.34 & 0.01 & 14.63 & 14.82 & 15.77 & seen\\ \hline

\end{tabular}}\end{center}
\end{table}

\begin{table}[htb]\begin{center}
\caption{\label{rel_AL1_3}Same as Table (\ref{rel_AL1}).}

\small{\begin{tabular}{|l||c|c|c||c||c||c||c|}
\hline

\multicolumn{8}{|c|}{ }\\							 
\multicolumn{8}{|c|}{$ ^1P_1\rightarrow \: ^1S_0 $ }\\
\multicolumn{8}{|c|}{ }\\ \hline
transition &  elec. & interfer. & magn. & tot(AL1)& tot(AP1) & tot(BD) & exp. \\ \hline 
$ b_1(1235)^+ \rightarrow \pi^+ \gamma $   & 148.68 &  &  & 148.68 & 152.76 & 142.14  &  227.20 $ \pm $ 58.60 \\ \hline

\multicolumn{8}{|c|}{ }\\
\multicolumn{8}{|c|}{$^3P_J \rightarrow \: ^1S_0$ }\\
\multicolumn{8}{|c|}{ }\\ \hline
transition &  elec. & interfer. & magn. & tot(AL1)& tot(AP1) & tot(BD) & exp. \\ \hline
$ a_1(1260)^+ \rightarrow \pi^+ \gamma $ & & & 179.53 & 179.53 & 229.90 & 151.63 & seen  \\ \hline
$ a_1(1260)^0 \rightarrow \pi^0 \gamma $ & & &  - & - &- &-  & seen  \\ \hline
$ a_2(1320)^+ \rightarrow \pi^+ \gamma $ & & & 142.01 & 142.01 & 179.27 & 120.87 & 299.60 $ \pm $ 65.71 \\ \hline

\multicolumn{8}{|c|}{ }\\
\multicolumn{8}{|c|}{$^1P_1 \rightarrow \: ^3S_1 $ }\\
\multicolumn{8}{|c|}{ }\\ \hline
transition &  elec. & interfer. & magn. & tot(AL1)& tot(AP1) & tot(BD) & exp. \\ \hline 
$ D_{s1}(2536)^{*+} \rightarrow D_s^{*+} \gamma$  & & & 10.97 & 10.97 & 11.99 & 9.15 & probably seen \\ \hline

\end{tabular}}\end{center}
\end{table}

\subsection{Mixing angles }

If the wave function is composed of two parts as in the $\eta$ mesons (flavor mixing), or in
the $\rho$ (isospin mixing with the $\omega$), some precisions are needed in the formalism.
In the case of $\eta$, the wave function can be written: 
$| \Psi_{\eta}> = | \Psi_{\eta_n}(n\bar{n}; I=0)> - | \Psi_{\eta_s}>$. In our study 
there is no instanton effect so we put by hand a mixing of 50\% between the two flavors.
That is we calculate separately the states $\tilde{\eta}_n$ and $\tilde{\eta}_s $ 
both normalised to one 
and we have $ | \Psi_{\eta_n}(n\bar{n}; I=0)>= \frac{\tilde{\eta}_n}{\sqrt{2}} $,
$ | \Psi_{\eta_s}>= \frac{\tilde{\eta}_s}{\sqrt{2}}$. A possible difficulty is 
that those states do not have the same mass in order to calculate the
phase space; nevertheless as we take the experimental value there is no difficulty.\\
From a general point of view we write  $| \Psi> = | \Psi_1> \pm | \Psi_2>$ where 
1 and 2 denote the two flavor (or isospin) components of the wave function. We have to calculate:
$$M(A\rightarrow B\gamma)= <A|{\bf M}|B> = (< A_1| \pm < A_2|){\bf M}( |B_1> \pm | B_2>)$$
so there could exist 4 components, and therefore some interferences. In the case of 
flavor mixing not all the terms will contribute due to flavor conservation. 

As we have said in section \ref{comments}, the $\phi$ could decay into the 
$\omega$, the $\rho$ and the $\pi$. This could be incorporated to our decay process
by two ways: an isospin mixing ($\omega$ and $\rho$) or a flavor mixing ($\phi$ and $\omega$). 
This study is done with the general formalism and a relativistic phase space. Because of the 
similarity of the results obtained via the three potentials, it is sufficient to 
perform the calculation with only one potential, here the AL1.

\subsubsection{Flavor mixing}
In this part, we investigate the mixing between the $\phi$ and the $\omega$.
Technically we use the prescription of section \ref{phi_om}. Now we have to find an 
appropriate value of the mixing angle $\theta_f$ 
and for that we rely on experimental data. A good candidate  
is the transition $\phi \rightarrow \pi^0 \gamma$ which is possible only
through a flavor mixing. Only the $n\bar{n}$ flavor part of 
the $\phi$ contributes to the decay. We find a small mixing of 
$\theta_f=4.5$ degrees. We could have choosen $\omega \rightarrow \eta \gamma$ 
to determine the angle value but the transition including a $\eta$ meson are
not very appropriate because of its flavor mixing which could generate some
interference terms. The $\omega \rightarrow \pi^0 \gamma$ transition is no
 more suited for this, even if only one term ($\omega_{n\bar{n}}$) contributes
because the value of a pure $n\bar{n}$ meson is smaller (459.30 keV) than the 
experimental value (714.85 keV) so including a $s\bar{s}$ part to the $\omega$
which will not contribute to the decay could only decrease the width.  

% {\small
 \begin{table}[ht]\begin{center}
\caption{\label{phiom} Decay widths obtained with a mixing angle of 
$\theta_f=4.479^o$ between the $\phi$ and $\omega$ mesons. Beyond the 
long wave length approximation, AL1 potential, relativistic phase space.
The mixing angle is obtained from the $\phi \rightarrow \pi^0 \gamma$.}
%\small

\begin{tabular}{|l||c|c||c|c|}\hline
\multicolumn{5}{|c|}{ } \\  
\multicolumn{5}{|c|}{$^3S_1 \rightarrow \: ^1S_0$} \\  
\multicolumn{5}{|c|}{ } \\  \hline
transition &  $k_0$ & $\frac{E_B}{m_A}$ & theo.  & exp. \\ \hline
$ \omega \rightarrow \pi^0 \gamma $    & 380 & 0.51 & 456.49  & 714.85 $\pm$ 42.74\\ \hline
$ \omega \rightarrow \eta \gamma  $    & 200 & 0.74 & 7.22  & 5.47 $ \pm $ 0.84 \\ \hline
$ \phi(1020) \rightarrow \pi^0 \gamma   $ &501 & 0.51 & fitted & 5.617 $\pm$ 0.447 \\ \hline
$ \phi(1020) \rightarrow \eta \gamma   $ & 363 & 0.64 & 35.88 & 55.82$\pm$2.73\\ \hline
$ \phi(1020) \rightarrow \eta'(958)\gamma$& 60 & 0.94 & 0.34 & 0.53 $\pm$ 0.31 \\ \hline
\multicolumn{5}{|c|}{ } \\  
\multicolumn{5}{|c|}{$^1S_0 \rightarrow \: ^3S_1$ } \\  
\multicolumn{5}{|c|}{ } \\  \hline
transition &  $k_0$ & $\frac{E_B}{m_A}$ & theo.  & exp. \\ \hline
$ \eta'(958) \rightarrow \omega \gamma $ & 159 & 0.83 & 8.58  & 6.11 $ \pm $ 0.78 \\ \hline
\multicolumn{5}{|c|}{ } \\  
\multicolumn{5}{|c|}{$^3P_J \rightarrow \: ^3S_1$ } \\  
\multicolumn{5}{|c|}{ } \\  \hline
transition &  $k_0$ & $\frac{E_B}{m_A}$ & theo.  & exp. \\ \hline
$ f_1(1285) \rightarrow \phi \gamma  $   & 236 & 0.82 & 0.48 &  17.76$ \pm $6.30   \\ \hline
\multicolumn{5}{|c|}{ } \\  
\multicolumn{5}{|c|}{$^3S_1 \rightarrow \:^3P_J $ } \\  
\multicolumn{5}{|c|}{ } \\  \hline
$\phi \rightarrow f_0(980) \gamma $ &39 &0.96 & 0.03 & 1.52$\pm$0.18 \\  \hline
$\phi \rightarrow a_0(980) \gamma $ &34 &0.97 & 0.23 & $<$ 22.29 \\  \hline

\end{tabular}\end{center}
\end{table}% }

The transitions modified in consequence are presented in table (\ref{phiom}).
First of all, we see that for the transitions already allowed, there is no 
superority taking into account mixing angles. Some transitions like the 
$\omega \rightarrow \eta \gamma$, $\phi(1020) \rightarrow \eta \gamma$ are  
deteriorated whereas the $\eta'(958) \rightarrow \omega \gamma$ is improved.
In fact the only improvement is the possibility to calculate 
transitions like the $f_1(1285) \rightarrow \phi \gamma$, 
$\phi \rightarrow f_0(980) \gamma$, $\phi \rightarrow a_0(980) \gamma$
which would be forbidden otherwise. 
Nevertheless the mixing is here to small to reproduce the experimental data.

\subsubsection{ \label{rho_om} Isospin mixing}
Here we consider an isospin mixing between the $\omega$ and the $\rho^0$.
This mixing do not appear for the charged $\rho^{\pm}$ because it is an 
$M_I=\pm 1$ meson. The angle $\theta_I$=8.9 degree is taken to fit the transition 
$\rho^0 \rightarrow \pi^0 \gamma$, and discriminates between charged and neutral
transitons. The results are presented in table (\ref{rhoom}). 
\begin{table}[htb]\begin{center}
\caption{ \label{rhoom} Decay widths obtained with a mixing angle of 
$\theta=8.88^o$ between the $\rho$ and $\omega$ mesons. Beyond the 
long wave length approximation, AL1 potential, relativistic phase space. 
the determination of the mixing angle is based on the transition 
$ \rho^0 \rightarrow \pi^0 \gamma$.}

\begin{tabular}{|l||c|c||c|c|}
\hline
\multicolumn{5}{|c|}{ } \\  
\multicolumn{5}{|c|}{$^3S_1 \rightarrow \: ^1S_0$} \\  
\multicolumn{5}{|c|}{ } \\  \hline
transition &  $k_0$ & $\frac{E_B}{m_A}$ & theo.  & exp. \\ \hline
$ \rho^0 \rightarrow \pi^0 \gamma $    & 373 & 0.52 & fitted  & 102.48$ \pm $ 25.69 \\ \hline
$ \rho^0 \rightarrow \eta \gamma $     & 190 & 0.75 & 51.57  & 36.18$ \pm $ 13.57 \\ \hline
$ \omega \rightarrow \pi^0 \gamma $    & 380 & 0.51 & 402.85  & 714.85 $\pm$ 42.74\\ \hline
$ \omega \rightarrow \eta \gamma  $    & 200 & 0.74 & 1.68  & 5.47 $ \pm $ 0.84 \\ \hline
\multicolumn{5}{|c|}{ } \\  
\multicolumn{5}{|c|}{$^1S_0 \rightarrow \: ^3S_1$ } \\  
\multicolumn{5}{|c|}{ } \\  \hline
transition &  $k_0$ & $\frac{E_B}{m_A}$ & theo.  & exp. \\ \hline
$ \eta'(958) \rightarrow \rho^0 \gamma $ & 170 & 0.82 & 121.99 & 61.31 $ \pm $ 5.51 \\ \hline
$ \eta'(958) \rightarrow \omega \gamma $ & 159 & 0.83 & 2.89  & 6.11 $ \pm $ 0.78 \\ \hline
\multicolumn{5}{|c|}{ } \\  
\multicolumn{5}{|c|}{$^3P_J \rightarrow \: ^3S_1$ } \\  
\multicolumn{5}{|c|}{ } \\  \hline
transition &  $k_0$ & $\frac{E_B}{m_A}$ & theo.  & exp. \\ \hline
$ f_1(1285) \rightarrow \rho^0 \gamma  $   & 410 & 0.68 & 1332.09  & 1296.00 $ \pm $ 295.20  \\ \hline 
\end{tabular}\end{center}
\end{table}

Considering the poor quality of the results (4 transitions are deteriorated and 2 
improved but not reproducing the experimental values), it is clear that we are 
missing something. This could be very well a false angle value, may be the 
$\rho^0 \rightarrow \pi^0 \gamma$ transition results of another process and should not
be use to determine $\theta_I$.

\section{Summary}
\indent
  
This work is a first review of the decay of a meson into another one plus a 
real photon. We have analyzed carefully the different parts of this elementary 
process. First of all we have presented different formalisms; the Long Wave Length 
Approximation, also called the static approximation, where the recoil term is 
neglected. We also investigated the Dipole Approximation, in which the spacial 
integrals are performed in the representation space with the presence of the factor
$(m_A-m_B)$; another step in this non relativistic approximation is the replacement 
of $(m_A-m_B)$ by the photon energy $E_0$. And finally we go beyong the Long Wave 
Length Approximation. The different formulae for the decay widths are presented in 
the appendix. In view of the results no formalism seems more suited than the others 
to describe the radiative decay, all of them giving the right order of magnitude, 
nevertheless the superiority of the general formalism comes from the fact that it 
allows the calculation of electric-magnetic interference terms and forbidden 
transitions in the LWLA and DA such as $^1S_0 \rightarrow\:^3P_J$.\\
Secondly, we study three phase spaces: relativistic, non relativistic and mixed. This 
last one uses a relativistic momentum but a non relativistic value of the ratio 
$\frac{E_B}{m_A}$ which allows us to show that the important point is a good 
description of the momentum. Anyhow we showed that a relativistic phase space
is always better and should be used in any circumstance.\\
Then we checked the importance of the wave function through the use of three 
potentials: AL1, AP1 and BD. Those calculations do not allow to show the 
superiority of one of them, the predicted values being of the same quality.
For our results to be analytical we expand our wave functions as a sum of N gaussian 
terms. We showed that N=3 is sufficient to obtain a convergence of the results but 
in order to be sure to treat the exact wave function, we used everywhere in our 
calculation N=5. We also showed that using exact wave functions is always
preferable as the one gaussian approximation that is sometimes of common use.
The difference could be as large as a 20\% effect.\\
Finally we incorporated some mixing angles in order to calculate some otherwise 
badly reproduced or even forbidden transitions such as $\phi\rightarrow \pi^0 \gamma$.
Those angles are of two kinds: isospin mixing between the $\rho$ and $\omega$ mesons
and flavor mixing between the $\phi$ and $\omega$ mesons. The results are not 
satisfactory. Except for a possible explanation of the important difference between
neutral and charged decay of the $\rho$ into pion, the isospin mixing deteriorate 
the quality of the predicted value whereas for the flavor mixing there is no 
important improvements.\\

Although a completely rigourous formalism gives an overall satisfactory agreement
with experimental data, we gave arguments that, in the framework that we
considered (NRQM and NR expression for the transition operator) some physics is still
absent. We think that one can explore two different directions : inclusion of
relativistic effects both in the wave functions and in the transition operator, and
introduction of form factors at the quark-photon vertex. Such works are under study. 

\section*{Appendix}
\subsection*{Long Wave Length Approximation}
In this part, we give the simplified expressions of (\ref{lte},\ref{ltm})
for the transitions of experimental interest, using wave functions expanded on 
gausssian terms as in (\ref{fop}).
\subsubsection*{Electric term}
$$\Gamma(^3S_1 \rightarrow\:^3\!P_J) = \frac{\pi\alpha}{48} \frac{E_B(E_0)}{m_a}\, E_0\,
\hat{J}^2 
\left(\frac{<e_1>}{m_1}- \frac{<e_2>}{m_2}\right)^2 
\left[ \sum_{i,j} \frac{c_i^A c_j^B}{D_{i,j}^{\frac{5}{2}}} \right]^2 $$

$$\Gamma(^3\!P_J \rightarrow\:^3\!S_1) = \frac{\pi\alpha}{16} \frac{E_B(E_0)}{m_a}\, E_0\,
\hat{J}^2 
\left(\frac{<e_1>}{m_1}- \frac{<e_2>}{m_2}\right)^2 
\left[ \sum_{i,j} \frac{c_i^A c_j^B}{D_{i,j}^{\frac{5}{2}}} \right]^2 $$

$$\Gamma(^1\!P_1 \rightarrow\:^1\!S_0) = \frac{\pi\alpha}{16} \frac{E_B(E_0)}{m_a}\, E_0\,
\hat{J}^2 
\left(\frac{<e_1>}{m_1}- \frac{<e_2>}{m_2}\right)^2 
\left[ \sum_{i,j} \frac{c_i^A c_j^B}{D_{i,j}^{\frac{5}{2}}} \right]^2 $$

\subsubsection*{Magnetic term}
$$\Gamma(^3\!S_1 \rightarrow\:^1\!S_0 ) = \frac{\pi\alpha}{48} \frac{E_B(E_0)}{m_a}\, E_0^3\,
\hat{J}^2 
\left(\frac{<e_1>}{m_1}- \frac{<e_2>}{m_2}\right)^2 
\left[ \sum_{i,j} \frac{c_i^A c_j^B}{D_{i,j}^{\frac{3}{2}}} \right]^2 $$

$$\Gamma(^1\!S_0 \rightarrow\:^3\!S_1) = \frac{\pi\alpha}{16} \frac{E_B(E_0)}{m_a}\, E_0^3\,
\hat{J}^2 
\left(\frac{<e_1>}{m_1}- \frac{<e_2>}{m_2}\right)^2 
\left[ \sum_{i,j} \frac{c_i^A c_j^B}{D_{i,j}^{\frac{3}{2}}} \right]^2 $$

$$\Gamma(^3\!S_1 \rightarrow\:^3\!S_1) = \frac{\pi\alpha}{24} \frac{E_B(E_0)}{m_a}\, E_0^3\,
\hat{J}^2 
\left(\frac{<e_1>}{m_1} + \frac{<e_2>}{m_2}\right)^2 
\left[ \sum_{i,j} \frac{c_i^A c_j^B}{D_{i,j}^{\frac{3}{2}}} \right]^2 $$
We see that for the last formula, this transition is null in the case of a meson
composed of only one flavor due to isospin term.

\subsubsection*{Forbidden transitions}
The following transitions are not allowed in the LWLA.

\begin{itemize}
\item $ ^1\!S_0 \rightarrow\:^3\!P_J$
\item $ ^3\!P_J  \rightarrow\:^1\!S_0$
\item $^1\!P_1 \rightarrow\:^3\!S_1$
\end{itemize}

\subsection*{General Case}
Here we give the simplified expression of the total exact width (\ref{larg}).
\subsubsection*{$L_b=0$}
We adopt the following notation for the decay width formulae:
$$ \Gamma_{A\rightarrow B\gamma} = \delta_{S_b,J_b}\ 2\alpha \frac{E_B(E_0)}{m_a}
\left[ \mathcal{EE} +\mathcal{EM} + \mathcal{MM}  \right]$$

\begin{itemize}
\item $ ^1\!S_0 \rightarrow\: ^1\!S_0$ 
$$ \Gamma_{A\rightarrow B\gamma} =0$$

\item $  ^1\!S_0 \rightarrow\:^3\!S_1 $
\begin{center}$\mathcal{EE}=0; \hspace{1cm} \mathcal{EM}=0$\end{center}
$$\mathcal{MM} = \frac{\pi}{32}\ E_0^3\ \left(\frac{<e_1>}{m_1}
\sum_{i,j} \frac{c_i^A c_j^Be^{-F_{ij}^{(1)}E_0^2}}{D_{i,j}^{\frac{3}{2}}}
-\frac{<e_2>}{m_2}\sum_{i,j} \frac{c_i^A c_j^Be^{-F^{(2)}_{ij}E_0^2}}{D_{i,j}^{\frac{3}{2}}}
 \right )^2$$ \\

\item $^3\!S_1 \rightarrow\:^1\!S_0$
\begin{center}$\mathcal{EE}=0; \hspace{1cm} \mathcal{EM}=0$\end{center}
$$ \mathcal{MM} = \frac{\pi}{96}\ E_0^3\ \left(\frac{<e_1>}{m_1}
\sum_{i,j} \frac{c_i^A c_j^Be^{-F_{ij}^{(1)}E_0^2}}{D_{i,j}^{\frac{3}{2}}}
-\frac{<e_2>}{m_2}\sum_{i,j} \frac{c_i^A c_j^Be^{-F^{(2)}_{ij}E_0^2}}{D_{i,j}^{\frac{3}{2}}}
 \right )^2 $$\\

\item $^3\!S_1 \rightarrow\:^3\!S_1$
\begin{center}$\mathcal{EE}=0; \hspace{1cm} \mathcal{EM}=0$\end{center}
$$ \mathcal{MM} = \frac{\pi}{48}\ E_0^3\ \left(\frac{<e_1>}{m_1}
\sum_{i,j} \frac{c_i^A c_j^Be^{-F_{ij}^{(1)}E_0^2}}{D_{i,j}^{\frac{3}{2}}}
+\frac{<e_2>}{m_2}\sum_{i,j} \frac{c_i^A c_j^Be^{-F^{(2)}_{ij}E_0^2}}{D_{i,j}^{\frac{3}{2}}}
 \right )^2 $$\\

\item $^1\!P_1 \rightarrow\:^1\!S_0$
\begin{center}$\mathcal{EM}=0; \hspace{1cm} \mathcal{MM}=0$\end{center}
$$ \mathcal{EE} = \frac{\pi}{32}\ E_0\ \left(\frac{<e_1>}{m_1}
\sum_{i,j} \frac{c_i^A c_j^Be^{-F_{ij}^{(1)}E_0^2}}{D_{i,j}^{\frac{5}{2}}}
-\frac{<e_2>}{m_2}\sum_{i,j} \frac{c_i^A c_j^Be^{-F^{(2)}_{ij}E_0^2}}{D_{i,j}^{\frac{5}{2}}}
 \right )^2 $$\\

\item $^1\!P_1 \rightarrow\:^3\!S_1$
\begin{center}$\mathcal{EE}=0; \hspace{1cm} \mathcal{EM}=0$\end{center}
$$ \mathcal{MM} = \frac{\pi}{32}\ E_0^5\ \left(\frac{<e_1>}{m_1}
\sum_{i,j} \frac{c_i^A c_j^B x^{(1)}_{ij}e^{-F_{ij}^{(1)}E_0^2}}{D_{i,j}^{\frac{3}{2}}}
+\frac{<e_2>}{m_2}\sum_{i,j} \frac{ c_i^A c_j^B x^{(2)}_{ij}e^{-F^{(2)}_{ij} E_0^{2}} }{D_{i,j}^{\frac{3}{2}}} \right )^2 $$\\

\item $^3\!P_J \rightarrow\:^1\!S_0$
\begin{center}$\mathcal{EE}=0; \hspace{1cm} \mathcal{EM}=0$
\begin{itemize} 
\item $J=0 \hspace{1cm} \mathcal{MM}=0$
\item $J=1 \hspace{1cm} \mathcal{MM}= \frac{\pi}{64}\ E_0^5\ \left(\frac{<e_1>}{m_1}
\sum_{i,j} \frac{c_i^A c_j^B x^{(1)}_{ij}e^{-F_{ij}^{(1)}E_0^2}}{D_{i,j}^{\frac{3}{2}}}
+\frac{<e_2>}{m_2}\sum_{i,j} \frac{c_i^A c_j^B x^{(2)}_{ij}e^{-F^{(2)}_{ij}E_0^2}}
{D_{i,j}^{\frac{3}{2}}} \right )^2  $
\item $J=2 \hspace{1cm} \mathcal{MM}=\frac{3\pi}{320}\ E_0^5\ \left(\frac{<e_1>}{m_1}
\sum_{i,j} \frac{c_i^A c_j^B x^{(1)}_{ij}e^{-F_{ij}^{(1)}E_0^2}}{D_{i,j}^{\frac{3}{2}}}
+\frac{<e_2>}{m_2}\sum_{i,j} \frac{c_i^A c_j^B x^{(2)}_{ij}e^{-F^{(2)}_{ij}E_0^2}}
{D_{i,j}^{\frac{3}{2}}} \right )^2  $
\end{itemize}\end{center}

\item $^3\!P_J \rightarrow\:^3\!S_1$
\begin{center}\begin{itemize} 

\item J=0 $$ \hspace{1cm}\mathcal{EE}=\frac{\pi}{32}\, E_0\mathcal{A}^2(E_0)
; \hspace{0.5cm}\mathcal{EM}=\frac{\pi}{16}\, E_0^3\mathcal{A}(E_0)\mathcal{B}(E_0); \hspace{0.5cm}
\mathcal{MM}=\frac{\pi}{32}\, E_0^5\mathcal{B}^2(E_0)$$

\item J=1 $$\hspace{1cm}\mathcal{EE}=\frac{\pi}{32}\, E_0\mathcal{A}^2(E_0)
; \hspace{0.5cm}\mathcal{EM}=\frac{\pi}{32}\, E_0^3\mathcal{A}(E_0)\mathcal{B}(E_0); \hspace{0.5cm}
\mathcal{MM}=\frac{\pi}{64}\, E_0^5\mathcal{B}^2(E_0)$$

\item J=2 $$\hspace{1cm}\mathcal{EE}=\frac{\pi}{32}\, E_0\mathcal{A}^2(E_0)
; \hspace{0.5cm}\mathcal{EM}=-\frac{\pi}{32}\, E_0^3\mathcal{A}(E_0)\mathcal{B}(E_0); \hspace{0.5cm}
\mathcal{MM}=\frac{7\pi}{320}\, E_0^5\mathcal{B}^2(E_0)$$

\end{itemize}\end{center}
with:
\begin{center}
$$\mathcal{A}(E_0) = \frac{<e_1>}{m_1}\sum_{i,j} \frac{c_i^A c_j^B e^{-F_{ij}^{(1)}E_0^2}}{D_{i,j}
^{\frac{5}{2}}}-\frac{<e_2>}{m_2}\sum_{i,j} \frac{c_i^A c_j^B e^{-F^{(2)}_{ij}E_0^2}}
{D_{i,j}^{\frac{5}{2}}} $$ 
$$\mathcal{B}(E_0) = \frac{<e_1>}{m_1}\sum_{i,j} \frac{c_i^A c_j^B x^{(1)}_{ij}e^{-F_{ij}^{(1)}E_0^2}}{D_{i,j}^{\frac{3}{2}}}
-\frac{<e_2>}{m_2}\sum_{i,j} \frac{c_i^A c_j^B x^{(2)}_{ij}e^{-F^{(2)}_{ij}E_0^2}}
{D_{i,j}^{\frac{3}{2}}} $$ 
\end{center}
\end{itemize}

\subsubsection*{$L_a=0$}
We adopt the following notation for the decay width formulae:
$$ \Gamma_{A\rightarrow B\gamma} = \delta_{S_a,J_a}\ 2\alpha \frac{E_B(E_0)}{m_a}
\left[ \mathcal{EE} +\mathcal{EM} + \mathcal{MM}  \right]$$

\begin{itemize}
\item $ ^1\!S_0 \rightarrow\: ^1\!P_1$ 
\begin{center}$\mathcal{EM}=0; \hspace{1cm} \mathcal{MM}=0$\end{center}
$$\mathcal{EE} = \frac{3\pi}{32}\ E_0\ \left(\frac{<e_1>}{m_1}
\sum_{i,j} \frac{c_i^A c_j^Be^{-F_{ij}^{(1)}E_0^2}}{D_{i,j}^{\frac{5}{2}}}
-\frac{<e_2>}{m_2}\sum_{i,j} \frac{c_i^A c_j^Be^{-F^{(2)}_{ij}E_0^2}}{D_{i,j}^{\frac{5}{2}}}
 \right )^2$$ \\

\item $^1\!S_0 \rightarrow\:^3\!P_J $
\begin{center}$\mathcal{EE}=0; \hspace{1cm} \mathcal{EM}=0$
\begin{itemize} 
\item $J=0 \hspace{1cm} \mathcal{MM}=0$
\item $J=1 \hspace{1cm} \mathcal{MM}= \frac{3\pi}{64}\ E_0^5\ \left(\frac{<e_1>}{m_1}
\sum_{i,j} \frac{c_i^A c_j^B z^{(1)}_{ij}e^{-F_{ij}^{(1)}E_0^2}}{D_{i,j}^{\frac{3}{2}}}
+\frac{<e_2>}{m_2}\sum_{i,j} \frac{c_i^A c_j^B z^{(2)}_{ij}e^{-F^{(2)}_{ij}E_0^2}}
{D_{i,j}^{\frac{3}{2}}} \right )^2  $
\item $J=2 \hspace{1cm} \mathcal{MM}=\frac{3\pi}{64}\ E_0^5\ \left(\frac{<e_1>}{m_1}
\sum_{i,j} \frac{c_i^A c_j^B z^{(1)}_{ij}e^{-F_{ij}^{(1)}E_0^2}}{D_{i,j}^{\frac{3}{2}}}
+\frac{<e_2>}{m_2}\sum_{i,j} \frac{c_i^A c_j^B z^{(2)}_{ij}e^{-F^{(2)}_{ij}E_0^2}}
{D_{i,j}^{\frac{3}{2}}} \right )^2  $
\end{itemize}\end{center}

\item $^3\!S_1 \rightarrow\:^1\!P_1$
\begin{center}$\mathcal{EE}=0; \hspace{1cm} \mathcal{EM}=0$\end{center}
$$\mathcal{MM} = \frac{\pi}{32}\ E_0^5\ \left(\frac{<e_1>}{m_1}
\sum_{i,j} \frac{c_i^A c_j^B z^{(1)}_{ij}e^{-F_{ij}^{(1)}E_0^2}}{D_{i,j}^{\frac{3}{2}}}
+\frac{<e_2>}{m_2}\sum_{i,j} \frac{c_i^A c_j^B z_{ij}^{(2)}e^{-F^{(2)}_{ij}E_0^2}}{D_{i,j}^{\frac{3}{2}}}
 \right )^2$$ \\

\item $^3\!S_1 \rightarrow\:^3\!P_J $
\begin{center}\begin{itemize} 

\item J=0 $$ \hspace{1cm}\mathcal{EE}=\frac{\pi}{96}\, E_0\mathcal{A}^2(E_0)
; \hspace{0.5cm}\mathcal{EM}=\frac{\pi}{48}\, E_0^3\mathcal{A}(E_0)\mathcal{B}(E_0); \hspace{0.5cm}
\mathcal{MM}=\frac{\pi}{96}\, E_0^5\mathcal{B}^2(E_0)$$

\item J=1 $$\hspace{1cm}\mathcal{EE}=\frac{\pi}{32}\, E_0\mathcal{A}^2(E_0)
; \hspace{0.5cm}\mathcal{EM}=\frac{\pi}{32}\, E_0^3\mathcal{A}(E_0)\mathcal{B}(E_0); \hspace{0.5cm}
\mathcal{MM}=\frac{\pi}{64}\, E_0^5\mathcal{B}^2(E_0)$$

\item J=2 $$\hspace{1cm}\mathcal{EE}=\frac{5\pi}{96}\, E_0\mathcal{A}^2(E_0)
; \hspace{0.5cm}\mathcal{EM}=-\frac{5\pi}{96}\, E_0^3\mathcal{A}(E_0)\mathcal{B}(E_0); \hspace{0.5cm}
\mathcal{MM}=\frac{7\pi}{192}\, E_0^5\mathcal{B}^2(E_0)$$

\end{itemize}\end{center}
with:
\begin{center}
$$\mathcal{A}(E_0) = \frac{<e_1>}{m_1}\sum_{i,j} \frac{c_i^A c_j^B e^{-F_{ij}^{(1)}E_0^2}}{D_{i,j}
^{\frac{5}{2}}}-\frac{<e_2>}{m_2}\sum_{i,j} \frac{c_i^A c_j^B e^{-F^{(2)}_{ij}E_0^2}}
{D_{i,j}^{\frac{5}{2}}} $$ 
$$\mathcal{B}(E_0) = \frac{<e_1>}{m_1}\sum_{i,j} \frac{c_i^A c_j^B z_{ij}^{(1)}e^{-F_{ij}^{(1)}E_0^2}}{D_{i,j}^{\frac{3}{2}}}
-\frac{<e_2>}{m_2}\sum_{i,j} \frac{c_i^A c_j^B z^{(2)}_{ij}e^{-F^{(2)}_{ij}E_0^2}}
{D_{i,j}^{\frac{3}{2}}} $$ 
\end{center}
\end{itemize}

\section{acknowledgments}
We are very grateful to Dr B. Desplanques for interesting discussions, and to L. Blanco
for drawing our attention to important aspects concerning this work.


\begin{thebibliography}{99}

\bibitem{Isg78} N. Isgur and G. Karl,
Phys. Rev. {\bf D18}, 4187 (1978); {\bf D19}, 2653 (1979); {\bf D20}, 1191 (1979).

\bibitem{Sem92} C. Semay and B. Silvestre-Brac,
Phys. Rev. {\bf D46}, 5177 (1992).

\bibitem{Sem97} C. Semay and B. Silvestre-Brac,
Nucl. Phys. {\bf A618}, 455 (1997); {\bf A647}, 72 (1999).

\bibitem{Sin70} P. Singer
 Phys. Rev. {\bf D1}, 86 (1970).

\bibitem{Bro77} L.M. Brown, P. Singer
 Phys. Rev. {\bf D15}, 3484 (1977)

\bibitem{Hay76} P. Hays, M.V.K. Ulehla,
 Phys. Rev. {\bf D13}, 1339 (1976).

\bibitem{Hac78} R.H Hackman, N.G. Deshpande, D.A. Dicus, V.L. Teplitz
 Phys. Rev. {\bf D18}, 2537 (1978).

\bibitem{Cha84} P.K. Chatley, C.P Singh, M.P. Khanna
Phys. Rev {\bf D29}, 96 (1984).

\bibitem{Odo81} P.J. O'Donnell,
Rev. Mod. Phys. {\bf 53}, 673 (1981)

\bibitem{Bar76} T. Barnes,
Phys. Lett. {\bf B63}, 65 (1976)

\bibitem{Eic78} E. Eichten, K. Gottfried, T. Kinoshita, K.D. Lane, J.M. Yan,
Phys. Rev. {\bf D17}, 3090 (1978); {\bf D21}, 203 (1980)

\bibitem{Mcc83} R. Mc Clary, N. Byers,
Phys. Rev. {\bf D28}, 1692 (1983)

\bibitem{Bar92} N. Barik, P.C. Dash, A.R Panda,
Phys. Rev. {\bf D46}, 3856 (1992)

\bibitem{Bar98} N. Barik, S. Kar, P.C. Dash,
Phys. Rev. {\bf D57}, 405 (1998)

\bibitem{God85} S. Godfrey, N. Isgur,
Phys. Rev. {\bf D32}, 189 (1985)

\bibitem{Ell93} El hassan El aaoud, Riazuddin
Phys. Rev. {\bf D47}, 1026 (1993)

\bibitem{Bha81} R.K. Bhaduri, L.E. Cohler, Y. Nogami,
Il Nuovo Cimento {\bf A65}, 376 (1981)


\bibitem{Sil93} B. Silvestre-Brac and C Semay, ISN {\bf 93-69}
(unpublished);
B. Silvestre-Brac, Few-Body Syst. {\bf 20}, 1 (1996);
C. Semay and B. Silvestre-Brac, Z. Phys. {\bf C61}, 271 (1994).

\bibitem{Pan97} A.R. Panda, K.C. Roy,
 Int. J. Mod. Phy. {\bf E6}, 121 (1997).

\end{thebibliography}
\end{document}